\documentclass{elsart}
\newcommand{\hs}{\hspace*{0.5cm}}

\newcommand{\be}{\begin{equation}}
\newcommand{\ee}{\end{equation}}
\newcommand{\bea}{\begin{eqnarray}}
\newcommand{\eea}{\end{eqnarray}}
\newcommand{\nn}{\nonumber}
\newcommand{\crn}{\nonumber \\}

\newcommand{\al}{\alpha}
\newcommand{\la}{\lambda}

\newcommand{\bc}{\begin{center}}
\newcommand{\ec}{\end{center}}

\newcommand {\ba}{\begin{array}}
\newcommand {\ea}{\end{array}}
\newcommand{\ben}{\begin{enumerate}}
\newcommand{\een}{\end{enumerate}}
\usepackage{axodraw,epsfig, graphicx}
\begin{document}
\runauthor{ Huong, Long}
\begin{frontmatter}
\title{Lepton-flavor violating decays of   neutral Higgs  to muon
and tauon in  supersymmetric economical 3-3-1 model}
\author[]{ P.T.Giang\thanksref{G}}
\author[]{L.T.Hue\thanksref{H1}}
\author[]{D.T.Huong\thanksref{H}}
\author[]{H.N.Long\thanksref{L}}

\address{Institute of Physics, VAST, 10 Dao Tan, Ba Dinh, Hanoi, Vietnam}
\thanks[G]{Email: ptgiang@grad.iop.vast.ac.vn}
\thanks[H1]{Email: lthue@iop.vast.ac.vn}
\thanks[H]{Email: dthuong@iop.vast.ac.vn}
\thanks[L]{Email: hnlong@iop.vast.ac.vn}

\begin{abstract}

We investigate  Lepton-Flavor Violating (LFV) decays of  Higgs to
muon-tau in the Supersymmetric Economical 3-3-1 (SUSYE331) model.
In the presence of flavor  mixing  in  sleptons $\{\tilde{\mu},
\tilde{\tau} \}$ and large values of $v/v'$, the ratio of $Br(H
\rightarrow \tau^+\mu^-)/Br(H \rightarrow \tau^+\tau^-)$  can
reach  non-negligible values $\mathcal{O}(10^{-3})$, as in many
known SUSY models.
 We predict  that for the Standard Model  Higgs boson,
the LHC may detect its decay
  to muon and tauon.
  We also investigate the asymmetry between left and right LFV values of
corrections and prove that the LFV effects are dominated by the
left FLV term, which is $\mathcal{O}(10^3)$ times larger than the
right LFV term in the limit of small values of
$|\mu_\rho|/m_{SUSY}$.
 The contributions of Higgs-mediated effects to the
decay $\tau \rightarrow \mu\mu\mu$ are also discussed.

\end{abstract}
\begin{keyword}
  Supersymmetric models, Decays of taus, Supersymmetric Higgs bosons
 \PACS 12.60.Jv \sep 13.35.Dx\sep 14.80.Da

\end{keyword}
\end{frontmatter}

\section{\label{intro}Introduction}
The experimental evidences of non-zero neutrino masses and mixing
\cite{pdg} have shown that the Standard Model (SM) of fundamental
particles and interactions must be extended.
 Among many extensions
of the SM  known today, the models based on gauge symmetry
$\mathrm{SU}(3)_C\otimes \mathrm{SU}(3)_L \otimes \mathrm{U}(1)_X$
(called 3-3-1 models) \cite{331m,331r} have interesting features.
The model requires that the number of fermion families $N$ be a
multiple of the quark color in order to cancel anomalies, which
suggests an interesting connection between the number of flavors
and the strong color group.
 If further one uses the
condition of QCD asymptotic freedom, which is valid only if the
number of families of quarks is to be less than five, it follows
that $N$ is equal to 3. In addition, the third quark generation
has to be different from the first two, so this leads to the
possible explanation of why top quark is uncharacteristically
heavy (see, for example,~\cite{longvan}).  The 3-3-1 models can
also provide a solution of electric charge quantization observed
in the nature \cite{ecq}.

In one of the 3-3-1  models  \cite{331r}   three
$\mathrm{SU}(3)_L$ lepton triplets are of the form $(\nu_l, l,
\nu_l^c)_L$, where $\nu_l^c$ is related to the right-handed
component of the neutrino field $\nu_l$ (a model with right-handed
neutrinos). The scalar sector of this model requires three Higgs
triplets, and  it is interesting to note that two Higgs triplets
has the same $\mathrm{U}(1)_X$ charge with two neutral components
at their top and bottom. Giving all neutral Higgs fields a vacuum
expectation value (VEV), we can remove one  Higgs triplet. Hence
the Higgs sector of the  obtained model becomes minimal, and it
has been called the economical 3-3-1 model \cite{ecn331r}.

The lepton-flavor is  absolutely conserved in the SM. Recently,
experiments on neutrino oscillations have proved that lepton
flavor is not conserved. It leads to motivation on a search for
the signals of lepton flavor violations (LFV) beyond the SM.
 Many versions of the extension of Minimal Supersymmetric Standard
 Model (MSSM) with large $\tan\beta$ have been investigated in Higgs LFV decay.
 The interesting here is there
 exists parameter space that  predicts the branching ratio of these types of
 decays  are very sizable, enough to be detected by present colliders
 such as CERN Large Hadron Collider (LHC) \cite{LHCAtlas} or International Linear
 Collider (ILC) \cite{ILC}. For example,  the SM \cite{thamkhao}
 predicted that  branching ratio of $H \rightarrow \mu\tau$ is  very suppressed.
  However, in the beyond   SM, this ratio can reach
   large values, more than $10^{-4}$. In particular, Refs.
 \cite{Anna1,Anna2} showed that, in the MSSM, $BR (H \rightarrow \mu^+\tau^-)
 \sim  10^{-4}$  if  $m_H/M_{SUSY} \sim 10^{-1} $ .  The Minimal
 Supersymmetric Neutrino Seesaw Models ($\nu$MSSM)  \cite{cruz} predicted  the branching
  ratio of heavy Higgs LFV  decay  is  of order $10^{-4}$ while that of
 light Higgs LFV decay is of  order  $ 10^{-8}$.  For more
 discussions in details
 about the Higgs LFV decay, readers
are referred to \cite{black} for general LFV framework, to
\cite{Koji} for two Higgs doublet models, to
\cite{black,Koji,Babu1,Babu,Guasch,Cannoni,Carena} for MSSM, and
$\nu$MSSM and to \cite{Goto} for little Higgs models (LTH).

The MSSM  has shown that in the limit of large
 $\tan\beta$,  the radiative corrections become non-negligible in many
 Higgs LFV decay  processes. For example, refs. \cite{Babu1,Guasch} showed that
 the ratio $R_{b/t} \equiv BR(H \rightarrow
 \bar{b}b)/ BR( H \rightarrow \bar{t}t)$ can be distinguished between the MSSM
 and non-supersymmetric models. The main reason  is that the
 Higgs boson couplings to down-type fermions
   receive a large corrections enhanced by $\tan\beta$. It leads to
 many interesting decay processes in
  quark sector such as $ b \rightarrow s \gamma$ \cite{Tom,Gino}. The large value
  of $\tan\beta$ also leads to  many interesting effects in the
  lepton sector, especially when the LFV source in sleptons  is
  included.

  Recently, the  Supersymmetric Economical 3-3-1 model (SUSYE331) has been constructed  \cite{Dong1}.
  Apart from interesting features  that  mentioned in refs. \cite{Dong1,Dong2,Dong3,Long2},
 the scalar sector is minimal, and therefore it has been called the economical.
 In a series of works  \cite{Dong1,Dong2,Dong3,Long2}, we have developed and proved that the
non-supersymmetric version \cite{ecn331r} and supersymmetric
version are consistent, realistic and very rich in physics. In the
previous work \cite{Dong2}, we skip the LFV source in the soft
sector. However,  the model predicts more interesting
phenomenology  if there exists
 LFV source in the soft breaking terms.  In this paper, we will
 concentrate  on LFV  Higgs decays  to  $\mu\tau$ with
the presence of misalignment of sleptons
$\{\tilde{\mu},\tilde{\tau}\}$ and their sneutrinos contained in
soft breaking terms. In SUSYE331 model, for generating fermion
masses as well as canceling anomaly, one needs four Higgs
triplets. In particular,  the "up" $\rho^0$ Higgs
    gives mass for neutrinos and the remain, "down"  $ \rho^{\prime
    0}$, gives mass for  charged leptons  \cite{Dong1,Dong3}
     and other Higgs give mass for
    quarks.{The ratio of VEVs,
    namely $\frac{<\rho^0>}{<\rho^{\prime0}>}$, is denoted by
    $ \tan\gamma$ which is similar to $\tan\beta$ in MSMS. Hence,the  $\rho^0$  and $ \rho^{\prime
    0}$ Higgs play very important roles if we consider effects of  radiative correction
    in lepton sector in the limit of large $\tan\gamma$. The corrections
     may cause many non-negligible effects, such as the
     correction of lepton mass, branching ratio of LFV Higgs
     decay...On the other hand, the model 331 is
     the extension of SM  based on
     extended gauge symmetry. Therefore, comparing to MSSM,
     the SUSYE331 model contains
      new gauge bosons and new Higgses  as well as their
     superpartners. Because of
     appearing of new particles, the
     number of diagrams contributing to LFV  Higgs decay in SUSYE331 model
     is predicted more than that in MSSM. It leads to
     LFV in Higgs decay effected in SUSYE331  may be larger than
     in MSSM. Hence, in this work, we  investigate the  flavor
     violating Higgs coupling
     in SUSYE331 model, specially we focus on the
    $\{\mu,\tau\}$  generations.

 Our work is arranged  as follows: In Section \ref{parcontent},  we review
 the particle content in SUSYE331 model. The analytic expressions of  the Higgs
 effective couplings are studied in Section \ref{effectiveIn}. In Section \ref{numerical}, we study
 a numerical estimation on decay $H\rightarrow
  \mu\tau$ at colliders and compare contribution from the left and right
  LFV radiative corrections into the mentioned decay. In this section, we  also consider the
  contributions  of Higgs exchange to branching ratio of  $\tau\rightarrow 3\mu$ decay.
In  the last section, we summarize our   main results.

\section{\label{parcontent}Particle content }

Let us give brief report on the particle content in SUSYE331 model
\cite{Dong1}.  The superfields in  the anomaly-free model are
given by
 \be
\widehat{L}_{a L}=\left(\widehat{\nu}_{a}, \widehat{l}_{a},
\widehat{\nu}^c_{a}\right)^T_{L} \sim (1,3,-1/3),\hs
  \widehat {l}^{c}_{a L} \sim (1,1,1), \hs a=1,2,3 \label{l2}
\end{equation}
 \be \widehat Q_{1L}= \left(\widehat { u}_1,\
                        \widehat {d}_1,\
                        \widehat {u}^\prime
 \right)^T_L \sim (3,3,1/3),\nn \ee
\be \widehat {u}^{c}_{1L},\ \widehat { u}^{ \prime c}_{L} \sim
(3^*,1,-2/3),\widehat {d}^{c}_{1L} \sim (3^*,1,1/3 ), \label{l5}
\ee
\begin{equation}
\begin{array}{ccc}
 \widehat{Q}_{\alpha L} = \left(\widehat
 {d}_{\alpha}, - \widehat{u}_{\alpha},
 \widehat{d^\prime}_{\alpha}\right)^T_{L}
  \sim (3,3^*,0), \hs \al=2,3, \label{l3}
\end{array}
\end{equation}
\begin{equation}
\widehat{u}^{c}_{\alpha L} \sim \left(3^*,1,-2/3 \right),\hs
\widehat{d}^{c}_{\alpha L},\ \widehat{d}^{\prime c}_{\alpha L}
\sim \left(3^*,1,1/3 \right),\label{l4}
\end{equation}
\be \widehat{\chi}= \left ( \widehat{\chi}^0_1, \widehat{\chi}^-,
\widehat{\chi}^0_2 \right)^T\sim (1,3,-1/3), \hs \widehat{\rho}=
\left (\widehat{\rho}^+_1, \widehat{\rho}^0,
\widehat{\rho}^+_2\right)^T \sim  (1,3,2/3), \label{l8} \ee
 \be \widehat{\chi}^\prime= \left
(\widehat{\chi}^{\prime o}_1, \widehat{\chi}^{\prime
+},\widehat{\chi}^{\prime o}_2 \right)^T\sim ( 1,3^*,1/3), \,
\widehat{\rho}^\prime = \left (\widehat{\rho}^{\prime -}_1,
  \widehat{\rho}^{\prime o},  \widehat{\rho}^{\prime -}_2
\right)^T\sim (1,3^*,-2/3). \label{l10} \ee

Here we use some new notations as
$\widehat{\psi}^c_L=(\widehat{\psi}_R)^c\equiv
\widehat{\psi}_R^{\dagger}$ and  exotic quarks are denoted by
usual quarks with prime-superscripts ($u'$ with the electric
charge $q_{u'}=2/3$ and $d'$ with $q_{d'}=-1/3$). The values in
each parenthesis show corresponding quantum numbers of the
$(\mathrm{SU}(3)_c, \mathrm{SU}(3)_L, \mathrm{U}(1)_X)$ symmetry.
In this model, the $ \mathrm{SU}(3)_L \otimes \mathrm{U}(1)_X$
gauge group is broken via two steps:
 \be \mathrm{SU}(3)_L \otimes
\mathrm{U}(1)_X \stackrel{w,w'}{\longrightarrow}\ \mathrm{SU}(2)_L
\otimes \mathrm{U}(1)_Y\stackrel{v,v',u,u'}{\longrightarrow}
\mathrm{U}(1)_{Q},\label{stages}\ee where the VEVs are defined by
\be
 \sqrt{2} \langle\chi\rangle^T = \left(u, 0, w\right),  \hs \hs \sqrt{2}
 \langle\chi^\prime\rangle^T = \left(u^\prime,  0,
 w^\prime\right)\ee \be
\sqrt{2}  \langle\rho\rangle^T = \left( 0, v, 0 \right), \hs \hs
\sqrt{2} \langle\rho^\prime\rangle^T = \left( 0, v^\prime,  0
\right).\ee
 The vector superfields $\widehat{V}_c$, $\widehat{V}$ and
$\widehat{V}^\prime$ containing the usual gauge bosons are
 given in \cite{Dong1,Dong3}. The supersymmetric
 model possessing a general Lagrangian is studied in \cite{Dong3}.
 In the following, only terms relevant to our calculations are displayed.

\section {\label{effectiveIn} Higgs-muon-tauon  effective interactions}
 In the SUSYE331 model \cite{Dong1,Dong3}, at
  the tree level, the  down-type leptons ($ e, \mu,
\tau$) only couple to the neutral Higgs $(\rho^{\prime 0})$
through the Yukawa interaction given by \bea
\mathcal{L}_{llH}=-\frac{\lambda_{1ab}}{3}(L_{aL}
l^c_{bL}\rho^{\prime 0}+\rm  H.c.). \label{leptonrho01} \eea
 In general case, $\la_{1ab} \neq 0$, the Lagrangian given in
(\ref{leptonrho01})  not only provides mass for the charged
leptons  but also gives the source of the lepton flavor mixing
 at  the tree level. It means that if the couplings
  $\lambda_{1ab}\neq 0$ with $(a\neq b)$, the LFV  processes, such as
 Higgs$\rightarrow\mu\tau$, must be existed.  In this case,
 our theory  predicts very large branching ratios of
 LFV processes which exceed to
  experimental results discussed in \cite{exp}.
 Hence, in the following calculation, we
 skip  the $\lambda_{1ab}$  with $ (a\neq
  b$) in (\ref{leptonrho01}) .

 Let us consider  another source of LFV which is
 caused by slepton mixing. More details of slepton mixing, one can find in
 Appendix \ref{eigenstate}. Because of slepton mixing, the leading
 effective interactions  of leptons with
 $\rho^o, \rho^{\prime o}$ Higgs  can appeare  at the one-loop order.
 In this paper, we will concentrate only on
  the couplings of Higgs with $\{\mu, \tau\}$ leptons.

 In order to consider the $\mu, \tau$ flavor mixing at
 the one loop level, first we rewrite the original Lagrangian (\ref{leptonrho01})
in terms of  two component spinor notations which are familiar to
those in literature, namely \bea -\mathcal{L}_{0\mu\tau} =
\left(Y_{\mu} \mu^c_L \mu_L+ Y_{\tau}
\tau^c_L\tau_L\right)\rho^{\prime0} + \mathrm{H.c.},
\label{massmutau1}\eea
 where $ Y_\mu\equiv \lambda_{122}/3, Y_\tau\equiv
 \lambda_{133}/3$.\\
 At the one-loop level, if we skip all of the terms which are
 proportional to $Y_{\mu}$ except terms
contributing to  mass of muon, then  Yukawa interactions
containing Higgs-lepton-lepton couplings can be divided into two
parts:
\begin{itemize}
\item The lepton-flavor
conversing (LFC) part given by
\bea -\Delta\mathcal{L}_{FC}&=&
\left(Y_{\mu} \Delta^{1\rho}_{\mu}+ Y_{\tau}
\Delta^{2\rho}_{\mu}\right) \mu^c_L \mu_L \rho^{0*}+ Y_{\tau}
\Delta^{\rho}_{\tau}\tau^c_L\tau_L\rho^{0*} \crn
 &+&\left(Y_{\mu} \Delta^{1\rho^{\prime}}_{\mu}+
Y_{\tau}  \Delta^{2\rho^{\prime}}_{\mu}\right) \mu^c_L \mu_L
\rho^{\prime0}+ Y_{\tau}
\Delta^{\rho^{\prime}}_{\tau}\tau^c_L\tau_L\rho^{\prime0} +
\mathrm{H.c.}, \label{massmutau1loop1}\eea
 \item  The lepton-flavor violating (LFV) part
 given as
\bea -\Delta\mathcal{L}_{FV}&=& Y_{\tau}  \left( \Delta^{\rho}_{L}
\tau^c_L \mu_L+ \Delta^{\rho}_{R}\mu^c_L\tau_L\right)\rho^{0*}
\crn
 &+&Y_{\tau} \left( \Delta^{\rho^{\prime}}_{L }
\tau^c_L \mu_L +
\Delta^{\rho^{\prime}}_{R}\mu^c_L\tau_L\right)\rho^{\prime0} + \rm
H.c.,
 \label{massmutau1loop2}\eea
 \end{itemize}
 where all of $\Delta^{1\rho}_{\mu}, \Delta^{2\rho}_{\mu}, \Delta^{1\rho'}_{\mu},
 \Delta^{2\rho'}_{\mu},\Delta^{\rho}_{\tau},
 \Delta^{\rho'}_{\tau},\Delta^{\rho}_L,\Delta^{\rho'}_L, \Delta^{\rho}_R$ and
 $\Delta^{\rho'}_R$ are the leading effective couplings. \\
From now on, for convenience, we use notation $\Delta$ to imply
any radiative correction of couplings appearing in
(\ref{massmutau1loop1}) and (\ref{massmutau1loop2}). Note that  $
\Delta$ is a dimensionless function of mass parameters and
$\Delta^{\rho}_\mu$,
 $\Delta^{\rho}_\tau$  are non-zero value
 even if we assume that there is no flavor mixing in slepton sector.
 We emphasize that  $\Delta^{\rho}_{\tau}$ is one of quantities affecting
 on  many observable quantities such as the ratio of branching ratios
 $ Br(H \rightarrow b\bar{b})/ BR (H \rightarrow \tau
 \bar{\tau})$.  The contribution of  $\Delta^{\rho}_{\tau}$ to that of branching ratios
 in the SUSY model is studied in  \cite{Babu1,Guasch}.
  The  diagrams which contribute to all of $\Delta$s are drawn in  Appendix \ref{Diagrams}.

  Now let us construct the total effective
 Lagrangian for Higgs, muon and tauon couplings
 in terms of physical eigenstates.
First we write down the whole Lagrangian coming from all of Eqs.
(\ref{massmutau1}), (\ref{massmutau1loop1}) and
(\ref{massmutau1loop2}) in the matrix form
 \bea  - \mathcal{L}
&=& Y_{\tau} \left(%
\begin{array}{cc}
   \mu^c_L & \tau^c_L \\
\end{array}%
\right) \mathcal{Y}_{l_1} \left(%
\begin{array}{c}
  \mu_L \\
  \tau_L \\
\end{array}%
\right)~ \rho^{\prime0}
+ Y_{\tau} \left(%
\begin{array}{cc}
   \mu^c_L & \tau^c_L \\
\end{array}%
\right) \mathcal{Y}_{l_2}\left(%
\begin{array}{c}
  \mu_L \\
  \tau_L \\
\end{array}%
\right) \rho^{0*} + \mathrm{ H.c.}, \label{LagrangLepton1}\eea
 where $\mathcal{Y}_{l_1}$ and $\mathcal{Y}_{l_2}$ are matrices
  defined by the following formulas:
\bea \mathcal{Y}_{l_1} = \left(%
\begin{array}{cc}
   \Delta^{o\rho'}_\mu & \Delta^{\rho'}_R \\
   \Delta^{\rho'}_L& 1+ \Delta^{\rho'}_\tau \\
\end{array}%
\right);  \hs  \hs  \mathcal{Y}_{l_2} = \left(%
\begin{array}{cc}
   \Delta^{o\rho}_{\mu}
  &
  \Delta^{\rho}_{R}
  \\
  \Delta^{\rho}_{L} & \Delta^{\rho}_{\tau} \\
\end{array}%
\right),  \label{LeptonMatrix1}\eea with $y \equiv Y_\mu/ Y_\tau$
, $\Delta^{o\rho}_\mu \equiv y\Delta^{1\rho}_\mu+
\Delta^{2\rho}_\mu$ and $ \Delta^{o\rho'}_\mu \equiv y+y
\Delta^{1\rho'}_\mu+ \Delta^{2\rho'}_\mu $.

 Because of loop corrections,  the mass matrix of the $\mu, \tau$
  in (\ref{LagrangLepton1})  is no longer diagonal.
 In order to find the physical  eigenstates of
 muon and tauon,  we expand the Higgs $\rho$ and
 $\rho^\prime$ around the
 vacuum expectation values. As a consequence, the mixing  mass matrix for the muon
  and tauon are
\bea -\mathcal{L}_{mass} &=&  Y_{\tau} v^\prime \left(%
\begin{array}{cc}
   \mu^c_L & \tau^c_L \\
\end{array}%
\right) \mathcal{Y}_{l} \left(%
\begin{array}{c}
  \mu_L \\
  \tau_L \\
\end{array}%
\right) + \mathrm{H.c.},\label{Lmass1} \eea
 where
\bea \mathcal{Y}_l &\equiv& \mathcal{Y}_{l_1}+ t_{\gamma}
\mathcal{Y}_{l_2} = ( 1+\Delta^{\rho'}_\tau+\Delta^{\rho}_\tau
t_\gamma ) \left(%
\begin{array}{cc}
  \epsilon_{\mu} &  \epsilon_R\\
  \epsilon_L & 1 \\
\end{array}%
\right) \crn
 &=& ( 1+ \Delta^{\rho'}_\tau +\Delta^{\rho}_\tau
t_\gamma ) Y_{\epsilon}, \label{massmatrix2} \eea with \bea
t_{\gamma}&\equiv&\tan\gamma = \frac{v}{v'}= \frac{\langle
\rho^0\rangle}{\langle \rho^{'0}\rangle}, \hs
\epsilon_{\mu}\equiv\frac{\Delta^{o\rho'}_{\mu}+
\Delta^{o\rho}_{\mu}t_{\gamma }}{ 1 +
\Delta^{\rho'}_\tau+\Delta^{\rho}_\tau t_\gamma}, \crn
 \epsilon_{L}&\equiv& \frac{ \Delta^{\rho'}_L+ \Delta^{\rho}_L t_{\gamma} }
 {1 + \Delta^{\rho'}_\tau+\Delta^{\rho}_\tau t_\gamma}
 ,\hs  \epsilon_{R}\equiv \frac{\Delta^{\rho'}_R+\Delta^{\rho}_R t_{\gamma}
 }{1 + \Delta^{\rho'}_\tau+\Delta^{\rho}_\tau t_\gamma}
  \label{Delta1}\eea
  and
  \bea
  Y_{\epsilon}=  \left(%
\begin{array}{cc}
  \epsilon_{\mu} &  \epsilon_R\\
  \epsilon_L & 1 \\
\end{array}%
\right)
 \label{huong1} \eea
It is easy to see that the mixing mass matrix of muon and tauon
given in (\ref{huong1}) is a general matrix.  Finding the mass
eigenvalues of left-right
 leptons is equivalent to finding  a
matrix $C$ satisfying:
 \bea C^{\dagger} \mathcal{Y}_{\epsilon}^{\dagger}
  \mathcal{Y}_{\epsilon} C
 &=& \left(%
\begin{array}{cc}
  y_\mu^2 & 0 \\
  0 & y^2_{\tau} \\
\end{array}%
\right) \equiv \mathcal{Y}^2_d.
 \label{massmatrix3}\eea
In our theory, the matrix $C$ can be found in a form
\bea C=\left(%
\begin{array}{cc}
  c_{\Lambda} & s_{\Lambda} \\
  -s_{\Lambda} & c_{\Lambda} \\
\end{array}%
\right),  \label{RoMatrix1}\eea
 where $ c_{\Lambda} \equiv \cos\Lambda, s_{\Lambda}
\equiv \sin\Lambda$ and $\Lambda$ is  the rotation angle given by
 \bea
t_{2\Lambda} \equiv \tan(2\Lambda) = \frac{2(\epsilon_\mu
\epsilon_R +\epsilon_L)}{1+\epsilon^2_R-(\epsilon^2_\mu+
\epsilon^2_L)}. \label{Gocquay1}\eea In addition, $ \mathcal{Y}_d
= \mathrm{diag } (y_{\mu}, ~y_{\tau}) $ in which  $(y_{\mu},
~y_{\tau})$ are defined as follows
  \bea y^2_\mu &=& r'-r s^2_{\Lambda}, \hs y^2_\tau = r'+ r
 c^2_{\Lambda},
 \label{eiMTM1}\eea
 where
 \bea
 r^2 & \equiv& 4(\epsilon_\mu
\epsilon_R +\epsilon_L)^2 + \left[ 1+\epsilon^2_R-(\epsilon^2_\mu+
\epsilon^2_L)\right]^2,\hs  r'\equiv
\epsilon^2_{\mu}+\epsilon^2_{L} ~.
  \label{length1}\eea
  Note that the mass eigenvalues of muon
   and tauon are proportional to $(y_\mu, y_\tau)$, namely
   \bea
  m_{\mu} = y_\mu Y_\tau v^\prime( 1+ \Delta^{\rho'}_\tau
  +\Delta^{\rho}_\tau), \hs  m_{\tau} = y_\tau Y_\tau v^\prime
  ( 1+ \Delta^{\rho'}_\tau +\Delta^{\rho}_\tau).
  \eea
 On the other hand, the mass eigenstates of leptons $(\mu, \tau)$ and $(\mu^c,
 \tau^c)$ are determined from two transformations
\bea  L^c &=& \left(%
\begin{array}{c}
  \mu^c \\
  \tau^c \\
\end{array}%
\right) = (U_l)^T\left( \begin{array}{c}
  \mu^c_L \\
  \tau^c_L \\
\end{array}%
\right) = U^T_l L^c_L,\crn
L &=& \left(%
\begin{array}{c}
  \mu \\
  \tau \\
\end{array}%
\right) = V_l\left( \begin{array}{c}
  \mu_L \\
  \tau_L \\
\end{array}%
\right) = V_l L_L,\label{Maeigenstate1}\eea where $U_l$ and $V_l$
have come from (\ref{massmatrix3}), namely
 \bea U^{\dagger}_l &=& \mathcal{Y}^{-1}_d C^{\dagger}
 \mathcal{Y}^{\dagger}_{\epsilon}= \left(%
\begin{array}{cc}
  \frac{1}{y_\mu} & 0 \\
  0 & \frac{1}{y_\tau} \\
\end{array}%
\right) \left(%
\begin{array}{cc}
  c_{\Lambda} & -s_{\Lambda} \\
  s_{\Lambda} & c_{\Lambda}\\
\end{array}%
\right) \left(%
\begin{array}{cc}
  \epsilon_\mu & \epsilon_L \\
  \epsilon_R & 1 \\
\end{array}%
\right), \crn
V_l^{\dagger}&=& C =  \left(%
\begin{array}{cc}
  c_{\Lambda} & s_{\Lambda} \\
  -s_{\Lambda} & c_{\Lambda}\\
\end{array}%
\right).\label{CheoMaTran2}\eea
 Next, we  replace  $\mathcal{Y}_{l_1}$ in Eq. (\ref{LagrangLepton1}) by
  a new form deduced from Eq.(\ref{massmatrix2})
  $$ \mathcal{Y}_{l_1} = ( 1+
 \Delta^{\rho'}_\tau  +\Delta^{\rho}_\tau
t_\gamma ) Y_{\epsilon} - \mathcal{Y}_{l_2} t_{\gamma}$$
 Now we have obtained a new expression of (\ref{LagrangLepton1}) as
 follows
 \bea  - \mathcal{L}
&=& Y_{\tau}( 1+
 \Delta^{\rho'}_\tau  +\Delta^{\rho}_\tau
t_\gamma ) \left(%
\begin{array}{cc}
   \mu^c_L & \tau^c_L \\
\end{array}%
\right) \mathcal{Y}_{\epsilon} \left(%
\begin{array}{c}
  \mu_L \\
  \tau_L \\
\end{array}%
\right)~ \rho^{\prime o} \crn
&+& Y_{\tau} \left(%
\begin{array}{cc}
   \mu^c_L & \tau^c_L \\
\end{array}%
\right) \mathcal{Y}_{l_2}\left(%
\begin{array}{c}
  \mu_L \\
  \tau_L \\
\end{array}%
\right) (\rho^{o*}- t_{\gamma} \rho^{\prime o}) + \mathrm{ H.c.},
\label{LagrangLepton2}\eea In the basis of mass eigenstates of the
muon and tauon
 given in Eq. (\ref{Maeigenstate1}), the  Lagrangian (\ref{LagrangLepton2})
   transforms  into
\bea - \mathcal{L}_{d} &=& Y_{\tau}( 1+
 \Delta^{\rho'}_\tau  + \Delta^{\rho}_\tau
t_\gamma ) L^{cT} ~\mathcal{Y}_d~ L \rho^{\prime o} \crn \crn &+&
Y_\tau L^{cT} (U^{\dagger}_l \mathcal{Y}_2 V_l^{\dagger})L
(\rho^{o*}- t_{\gamma}\rho^{\prime o}) + \mathrm{H.c.}
\label{LagrangLepton3}\eea It is needed to emphasize that the
first term in Eq. (\ref{LagrangLepton3})
  generates only masses for muon and tauon while
 the second  creates masses as well as give rise to the lepton flavor mixing.
 Sources of flavor mixing  are two off-diagonal elements
  of   the matrix  $(U^{\dagger}_l \mathcal{Y}_2
 V_l^{\dagger})$ :
\bea  \left(U^{\dagger}_l \mathcal{Y}_{l_2}
V^{\dagger}\right)_{12} &=& \frac{c^2_{\Lambda} \Delta^{\rho}_{R}
\epsilon_\mu}{y_\mu} + \frac{(c^2_{\Lambda} \epsilon_L
-c_{\Lambda}s_{\Lambda})  \Delta^{\rho}_{\tau} }{y_\mu} \crn &+&
\frac{ c_{\Lambda}s_{\Lambda} (\Delta^{\rho}_{L} \epsilon_L +
\Delta^{o\rho}_{\mu}\epsilon_\mu -\Delta^{\rho}_{R}\epsilon_R)-
s^2_{\Lambda}(\Delta^{\rho}_L +\epsilon_R \Delta^{\rho}_R)
}{y_\mu}, \label{ViPhamL5}\\
 \left(U_l^{\dagger} \mathcal{Y}_{l_2} V^{\dagger}\right)_{21}&=&
\frac{c^2_{\Lambda} \Delta^{\rho}_L}{y_\tau} + \frac{c^2_{\Lambda}
\Delta^{o\rho}_{\mu} \epsilon_R -  c_{\Lambda}s_{\Lambda}
\Delta^{\rho}_{\tau} }{y_\tau}\crn
 &+&\frac{ c_{\Lambda}s_{\Lambda} (
\Delta^{\rho}_{L} \epsilon_L + \Delta^{o\rho}_{\mu}\epsilon_\mu
-\Delta^{\rho}_{R}\epsilon_R)- s^2_{\Lambda}(\Delta^{\rho}_\tau
\epsilon_L +\epsilon_\mu \Delta^{\rho}_R) }{y_\tau}
\label{ViPhamL6}\eea
  In the further calculations, we consider a case of
  $ (t_{\gamma} \Delta) \ll  1 $
  but  large enough (as investigated in MSSM) to cause many
  interesting effects, and we will comment more details after some
  numerical calculations. On the other hand, the
  rotation angle given in Eq. (\ref{Gocquay1}) is very small, so
  we can set $ c_{\Lambda}\simeq
 1,s_{\Lambda}\simeq \Lambda $.  As a result,
  Eqs.  (\ref{eiMTM1}), (\ref{ViPhamL5})
  and (\ref{ViPhamL6}) can be  presented as very simple formulas:
 \bea  y_{\mu} &\simeq& \epsilon_\mu, \hs y_{\tau} \simeq 1, \crn
  \left(U^{\dagger}_l \mathcal{Y}_{l_2}
V^{\dagger}\right)_{12} &\simeq& \Delta^{\rho}_R, \hs
\left(U^{\dagger}_l \mathcal{Y}_{l_2}
V^{\dagger}\right)_{21}\simeq \Delta^{\rho}_L,
\label{Deltasmall}\eea
 and the above LFV Lagrangian also appears in a simple form:
 \bea - \mathcal{L}_{FV}  &\simeq& Y_\tau(\Delta^{\rho}_R \mu^c \tau
 + \Delta^{\rho}_L
 \tau^c\mu ) (\rho^{0*}-t_{\gamma} \rho^{\prime 0})
 + \mathrm{H.c.}.\label{FCNC2}\eea
Finally,  in the mass-eigenstate  basis for both lepton and Higgs,
we obtain the effective LFV Lagrangian: \bea - \mathcal{L}_{FV}
&\simeq& \sqrt{2} Y_\tau (\Delta^{\rho}_R \mu^c \tau +
\Delta^{\rho}_L
 \tau^c\mu ) \left( s_{\alpha} s_{\gamma}
 \phi_{S_{a36}} - c_{\alpha} s_{\gamma} \varphi_{S_{a36}}\right) +
 \mathrm{H.c.},
 \label{FCNC3}\eea
where $\varphi_{S_{a36}}$ and  $\phi_{S_{a36}}$ are
 the Higgs mass eigenstates generated from the mixing of two original Higgs bosons
 $\rho^0$ and
$\rho^{\prime 0}$. The expressions of the Higgs mass eigenstates
 were introduced in \cite{Dong3}.
They are summarized in  the Appendix \ref{eigenstate}. The
emphasis here is that  in  the  general supersymmetric model
 there exist both the leading interactions of the muon, tauon with
neutral scalar and pseudo scalar Higgs.  However, the SUSYE331
model contains only interactions among muon, tauon and scalar
Higgs.

The effective couplings  given in (\ref{FCNC3}) are widely
investigated for many LFV low-energy processes, specially in the
MSSM \cite{Anna1,Cannoni,paradisi1}.
  In this paper we first concentrate on some simple
 aspects  of LFV in the SUSYE331 model. In particular, we are going to
  consider the LFV in
 decays of the scalar
  Higgs, i.e. $\Phi^0 \rightarrow \tau^{\pm}
 \mu^{\mp}$, where $\Phi^0 = \varphi_{S_{a36}}$ or
 $\phi_{S_{a36}}$. First,  we start with studying the branching
 ratios
  of neutral Higgs decay into muon and tauon. The SUSYE331 model
  predicts that the formula of these
   branching ratios is
\bea  BR(\Phi^0 \rightarrow \tau^+
 \mu^-) &=& BR(\Phi^0 \rightarrow \tau^-
 \mu^+) \crn
 &=& 2(1+ \tan^2 \gamma) \left(\mid\Delta^{\rho}_L\mid^2
 + \mid\Delta^{\rho}_R\mid^2
 \right) ~BR(\Phi^0 \rightarrow \tau^+
 \tau^- ). \crn \label{Br1}\eea
 This result is similar to that one given in
 \cite{Anna1}, except the absence of angle of mixing
  among Higgses. In the limit of appropriately  large $\tan\gamma$,
 the effects of LFV in the Higgs decay processes is not to
 be ignored. Hence,  our theoretical
 prediction is not much different from  that of previous results
 given in  \cite{Anna1,Anna2,Babu1,Babu,paradisi1}.
For details, we will study some numerical calculations for the
branching ratios indicated by Eq.(\ref{Br1}).  In our paper, we
use the assumption for slepton mixing presented in Appendix
\ref{eigenstate}. The diagrams giving contributions to
$\Delta^{\rho}_R$ and $\Delta^{\rho}_L$ are shown in
Fig.\ref{rho}.  The relevant vertices to our calculation
 are presented in Appendix \ref{Lagrangian1}.\\


\begin{figure}[t]
\vspace{0.4cm}
\begin{center}
\begin{picture}(120,80)(-60,-40)
\ArrowLine(-50,0)(-34,0) \ArrowLine(-17,0)(-34,0)
\ArrowLine(-17,0)(0,0) \ArrowLine(17,0)(0,0)
\ArrowLine(17,0)(34,0) \ArrowLine(50,0)(34,0)
\DashArrowArcn(0,-20)(40,150,30){4} \DashArrowLine(0,0)(0,-20){3}
\Text(-50,-10)[]{$\mu$} \Text(50,-8)[]{$\tau^c$}
\Text(27,-9)[]{\small $\tilde{\rho}^{\prime0}$}
\Text(11,-9)[]{\small $\tilde{\rho}^0$}
\Text(-17,-8)[]{\small$\lambda_B$}
 \Text(0,-28)[]{\small
$\small \rho^{0*}$} \Text(0,30)[]{$\tilde{l}_{L_\alpha}$}
\Text(-50,40)[]{$(a)$}
\end{picture}
\hglue 2.5cm
\begin{picture}(120,80)(-60,-40)
\ArrowLine(-50,0)(-34,0) \ArrowLine(-17,0)(-34,0)
\ArrowLine(-17,0)(0,0) \ArrowLine(17,0)(0,0)
\ArrowLine(17,0)(34,0) \ArrowLine(50,0)(34,0)
\DashArrowArcn(0,-20)(40,150,30){4} \DashArrowLine(0,0)(0,-20){3}
\Text(-50,-10)[]{$\mu$} \Text(50,-8)[]{$\tau^c$}
\Text(27,-9)[]{\small $\tilde{\rho}^{\prime0}$}
\Text(11,-9)[]{\small $\tilde{\rho}^0$}
\Text(-17,-8)[]{\small$\lambda^3_A$}
\Text(-20,-20)[]{\small$\lambda^8_A$} \Text(0,-28)[]{\small
$\rho^{0*}$} \Text(0,30)[]{$\tilde{l}_{L_\alpha}$}
\Text(-50,40)[]{$(b)$}
\end{picture}
\end{center}
\begin{center}
\begin{picture}(120,80)(-60,-40)
\ArrowLine(-50,0)(-34,0) \ArrowLine(-17,0)(-34,0)
\ArrowLine(-17,0)(0,0) \ArrowLine(17,0)(0,0)
\ArrowLine(17,0)(34,0) \ArrowLine(50,0)(34,0)
\DashArrowArcn(0,-20)(40,150,30){4} \DashArrowLine(0,0)(0,-20){3}
\Text(-50,-10)[]{$\mu$} \Text(50,-8)[]{$\tau^c$}
\Text(29,-9)[]{\small $\tilde{\rho}_1^{\prime-}$}
\Text(11,-9)[]{\small $\tilde{\rho}_1^{+}$} \Text(-9,-7)[]{\small
$\tilde{W}^-$} \Text(-27,-7)[]{\small $\tilde{W}^+$}
\Text(0,-28)[]{\small $\rho^{0*}$}
\Text(0,30)[]{$\tilde{\nu}_{L\alpha}$} \Text(-50,40)[]{$(c)$}
\end{picture}
\hglue 2.5cm
\begin{picture}(120,80)(-60,-40)
\ArrowLine(-50,0)(-34,0) \ArrowLine(-17,0)(-34,0)
\ArrowLine(-17,0)(0,0) \ArrowLine(17,0)(0,0)
\ArrowLine(17,0)(34,0) \ArrowLine(50,0)(34,0)
\DashArrowArcn(0,-20)(40,150,30){4} \DashArrowLine(0,0)(0,-20){3}
\Text(-50,-10)[]{$\mu$} \Text(50,-8)[]{$\tau^c$}
\Text(29,-9)[]{\small $\tilde{\rho}_2^{\prime-}$}
\Text(11,-9)[]{\small $\tilde{\rho}_2^{+}$} \Text(-9,-7)[]{\small
$\tilde{Y}^-$} \Text(-27,-7)[]{\small $\tilde{Y}^+$}
\Text(0,-28)[]{\small $\rho^{0*}$}
\Text(0,30)[]{$\tilde{\nu}_{R\alpha}$} \Text(-50,40)[]{$(d)$}
\end{picture}
\end{center}
\begin{center}
\begin{picture}(120,80)(-60,-40)
\ArrowLine(-50,0)(-34,0) \ArrowLine(0,0)(-34,0)
\ArrowLine(0,0)(34,0) \ArrowLine(50,0)(34,0)
\DashArrowArcn(0,-20)(40,150,90){4}
\DashArrowArcn(0,-20)(40,90,30){4} \DashArrowLine(0,20)(0,40){3}
\Text(-50,-10)[]{$\mu$} \Text(50,-8)[]{$\tau^c$}
\Text(15,-8)[]{\small $\tilde{\rho}^{\prime-}_2$}
\Text(-15,-8)[]{\small $\tilde{\rho}^{+}_2$} \Text(10,35)[]{\small
$\rho^{0*}$} \Text(-34,20)[]{$\tilde{\nu}_{L_\alpha}$}
\Text(34,20)[]{$\tilde{\nu}_{R_\beta}$} \Text(-50,40)[]{$(e)$}
\end{picture}
\hglue 2.5cm
\begin{picture}(120,80)(-60,-40)
\ArrowLine(-50,0)(-34,0) \ArrowLine(0,0)(-34,0)
\ArrowLine(0,0)(34,0) \ArrowLine(50,0)(34,0)
\DashArrowArcn(0,-20)(40,150,90){4}
\DashArrowArcn(0,-20)(40,90,30){4} \DashArrowLine(0,20)(0,40){3}
\Text(-50,-10)[]{$\mu$} \Text(50,-8)[]{$\tau^c$}
\Text(15,-8)[]{\small $\tilde{\rho}^{\prime-}_1$}
\Text(-15,-8)[]{\small $\tilde{\rho}^{+}_1$} \Text(10,35)[]{\small
$\rho^{0*}$} \Text(-34,20)[]{$\tilde{\nu}_{R_\alpha}$}
\Text(34,20)[]{$\tilde{\nu}_{L_\beta}$} \Text(-50,40)[]{$(f)$}
\end{picture}
\end{center}
\begin{center}
%
\begin{picture}(120,80)(-60,-40)
\ArrowLine(-50,0)(-34,0) \ArrowLine(-17,0)(-34,0)
\ArrowLine(-17,0)(0,0) \ArrowLine(17,0)(0,0)
\ArrowLine(17,0)(34,0) \ArrowLine(50,0)(34,0)
\DashArrowArcn(0,-20)(40,150,30){4} \DashArrowLine(0,0)(0,-20){3}
\Text(-50,-10)[]{$\tau$} \Text(50,-8)[]{$\mu^c$}
\Text(-27,-9)[]{\small $\tilde{\rho}^{\prime0}$}
\Text(-11,-9)[]{\small $\tilde{\rho}^0$} \Text(17,-8)[]{\small
$\lambda_B$} \Text(0,-28)[]{\small $\rho^{0*}$}
\Text(0,30)[]{$\tilde{l}_{R_\alpha}$} \Text(-50,40)[]{$(i)$}
\end{picture}
\end{center}
\begin{center}
\begin{picture}(120,80)(-60,-40)
\ArrowLine(-50,0)(-34,0) \ArrowLine(0,0)(-34,0)
\ArrowLine(0,0)(34,0) \ArrowLine(50,0)(34,0)
\DashArrowArcn(0,-20)(40,150,90){4}
\DashArrowArcn(0,-20)(40,90,30){4} \DashArrowLine(0,20)(0,40){3}
\Text(-50,-10)[]{$\mu$} \Text(50,-8)[]{$\tau^c$}
\Text(0,-8)[]{\small $\lambda_B$} \Text(10,35)[]{\small
$\rho^{*0}$} \Text(-34,20)[]{$\tilde{l}_{L_\alpha}$}
\Text(34,20)[]{$\tilde{l}_{R_\beta}$} \Text(-50,40)[]{$(k)$}
\end{picture}
\begin{picture}(120,80)(-60,-40)
\ArrowLine(-50,0)(-34,0) \ArrowLine(0,0)(-34,0)
\ArrowLine(0,0)(34,0) \ArrowLine(50,0)(34,0)
\DashArrowArcn(0,-20)(40,150,90){4}
\DashArrowArcn(0,-20)(40,90,30){4} \DashArrowLine(0,20)(0,40){3}
\Text(-50,-10)[]{$\tau$} \Text(50,-8)[]{$\mu^c$}
\Text(0,-8)[]{\small $\lambda_B$} \Text(10,35)[]{\small
$\rho^{*0}$} \Text(-34,20)[]{$\tilde{l}_{L_\alpha}$}
\Text(34,20)[]{$\tilde{l}_{R_\beta}$} \Text(-50,40)[]{$(l)$}
\end{picture}

\end{center}

\vspace{-0.3cm} \caption{\small Diagrams  contributing  to
$\Delta^{\rho}_L$ [$(a),(b), (c),(d), (e), (f), (k)$] and
$\Delta^{\rho}_R$ [$(i), (l)$].} \label{rho}
\end{figure}
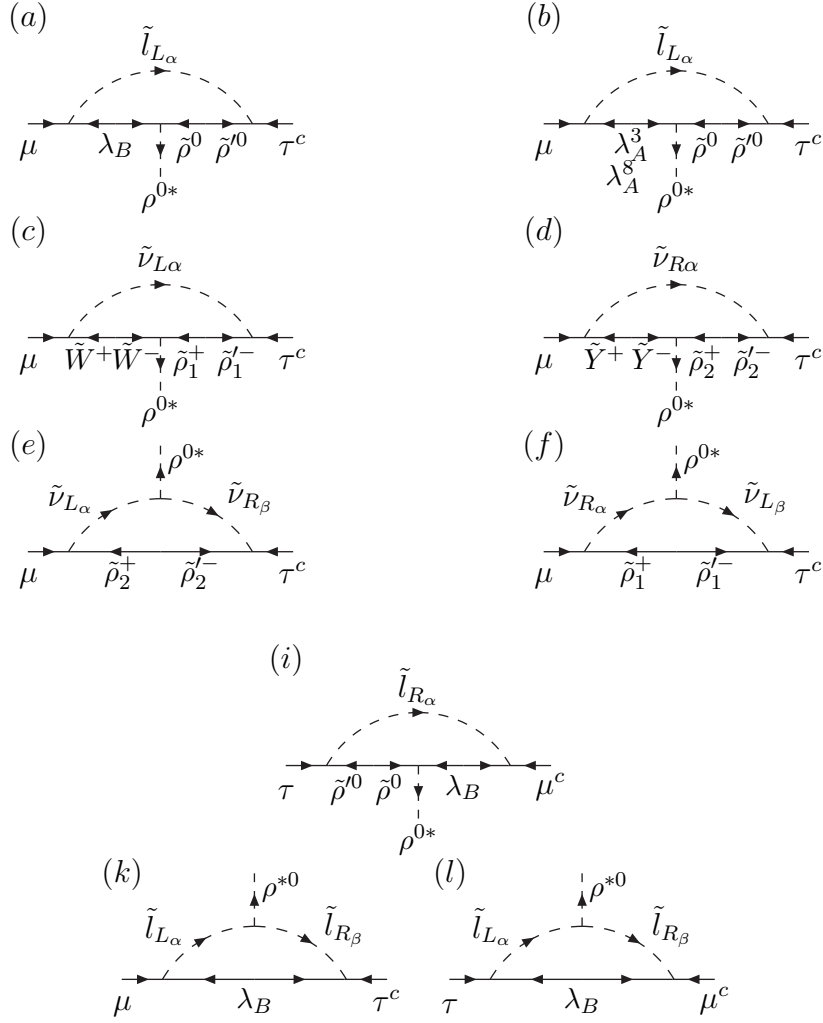

Using Feynman rules,  we can obtain the expression $\Delta_L^\rho$
from the diagrams in Fig.\ref{rho}, namely:
 \bea \Delta^{\rho}_L &=& \Delta^{\rho }_{La} +\Delta^{\rho
 }_{Lb} + \Delta^{\rho }_{Lc}+ \Delta^{\rho}_{Ld} + \Delta^{\rho}_{Le}+
  \Delta^{\rho}_{Lf} +\Delta^{\rho}_{Lk},
 \label{deltarho1}\eea
 where
\bea \Delta^{\rho}_{La} &=& \frac{g^{\prime 2}}{216 \pi^2}
\mu_{\rho} m' c_L s_L \left[ I_3(m^{\prime 2}, \mu^2_{\rho},
\tilde{m}^2_{L_2})- I_3 (m^{\prime 2}, \mu^2_{\rho},
\tilde{m}^2_{L_3}) \right], \crn
 \Delta^{\rho}_{Lb} &=& -\frac{ g^{2}}{24\pi^2}
\mu_{\rho} m_{\lambda} c_L s_L \left[ I_3(m^{2}_{\lambda},
\mu^2_{\rho}, \tilde{m}^2_{L_2})- I_3 (m^{2}_{\lambda},
\mu^2_{\rho}, \tilde{m}^2_{L_3}) \right],\crn
 \Delta^{\rho}_{Lc} &=& -\frac{g^{2}}{16 \pi^2}
\mu_{\rho} m_{\lambda} c_{\nu_L} s_{\nu_L} \left[
I_3(m^{2}_{\lambda}, \mu^2_{\rho}, \tilde{m}^2_{\nu_{L2}})- I_3
(m^{2}_{\lambda}, \mu^2_{\rho}, \tilde{m}^2_{\nu_{L3}})
\right],\crn
  \Delta^{\rho}_{Ld} &=& -\frac{g^{2}}{16 \pi^2}
\mu_{\rho} m_{\lambda} c_{\nu_R} s_{\nu_R} \left[
I_3(m^{2}_{\lambda}, \mu^2_{\rho}, \tilde{m}^2_{\nu_{R2}})- I_3
(m^{2}_{\lambda}, \mu^2_{\rho}, \tilde{m}^2_{\nu_{R3}}) \right],
\crn
  \Delta^{\rho}_{Le} &=&
  \frac{(Y_{\nu_{\mu\tau}})
  h_{\mu\tau}-h_{\tau\mu} \mu_{\rho}}{8
  \pi^2}\crn
 &\times& \left[s_{\nu_{(L-R)}} \left(s_{\nu_L} s_{\nu_R}
 I_3(\mu^2_{\rho},\tilde{m}^2_{\nu_{L2}},
 \tilde{m}^2_{\nu_{R2}})+  c_{\nu_L} c_{\nu_R}
 I_3(\mu^2_{\rho},\tilde{m}^2_{\nu_{L3}},
 \tilde{m}^2_{\nu_{R3}})  \right)\right.\crn
&+&  \left. c_{\nu_{(L-R)}} \left(s_{\nu_R} c_{\nu_L}
I_3(\mu^2_{\rho},\tilde{m}^2_{\nu_{L2}},
 \tilde{m}^2_{\nu_{R2}})-  s_{\nu_L} c_{\nu_R}
 I_3(\mu^2_{\rho},\tilde{m}^2_{\nu_{L2}},
 \tilde{m}^2_{\nu_{R3}})  \right) \right],\crn
  \Delta^{\rho}_{Lf} &=&-\Delta^{\rho }_{Le},\crn
\Delta^{\rho}_{Lk} &=& \frac{g^{\prime 2}}{288 \pi^2} \mu_{\rho}
m's_L c_L \left[s^2_R \left( I_3(m^{\prime 2},
\tilde{m}^2_{L_2},\tilde{m}^2_{R_2})- I_3(m^{\prime 2},
\tilde{m}^2_{L_3},\tilde{m}^2_{R_2} ) \right)\right.\crn &+&
\left. c^2_R \left( I_3(m^{\prime 2},
\tilde{m}^2_{L_2},\tilde{m}^2_{R_3})- I_3(m^{\prime 2},
\tilde{m}^2_{L_3},\tilde{m}^2_{R_3} )\right) \right].
    \label{deltarho2}\eea
Also,  the
 $\Delta^{\rho}_{R}$ receives contributions from  two
 diagrams ($i$) and ($l$) of Fig.\ref{rho} too,
\bea \Delta^{\rho}_{R} &=& \Delta^{\rho}_{Ri}+\Delta^{\rho}_{Rl}
,\label{deltarho3}\eea
 where
\bea \Delta^{\rho}_{Ri} &=& - \frac{g^{\prime 2}}{72 \pi^2}
\mu_{\rho} m' c_R s_R \left[ I_3(m^{\prime 2}, \mu^2_{\rho},
\tilde{m}^2_{R_2})- I_3 (m^{\prime 2}, \mu^2_{\rho},
\tilde{m}^2_{R_3}) \right],\crn \Delta^{\rho}_{Rl} &=&
\frac{g^{\prime 2}}{288 \pi^2} \mu_{\rho} m's_R c_R \left[s^2_L
\left( I_3(m^{\prime 2}, \tilde{m}^2_{L_2},\tilde{m}^2_{R_2})-
I_3(m^{\prime 2}, \tilde{m}^2_{L_2},\tilde{m}^2_{R_3} )
\right)\right.\crn &+& \left. c^2_L \left( I_3(m^{\prime 2},
\tilde{m}^2_{L_3},\tilde{m}^2_{R_2})- I_3(m^{\prime 2},
\tilde{m}^2_{L_3},\tilde{m}^2_{R_3} )\right) \right]
 \label{deltarho4}\eea
Here we have used some new notations
 \bea  s_{\nu_{(L-R)}} \equiv s_{\nu_L} c_{\nu_R} -s_{\nu_R} c_{\nu_L},
  \hs  c_{\nu_{(L-R)}} \equiv s_{\nu_L} s_{\nu_R} + c_{\nu_L}
 c_{\nu_R}, \label{subtracAngle1}\eea
 where $s_L$, $c_L$ and $s_R$, $c_R$ are deduced from  mixing angles for left and
 right handed sleptons (for details, see  Appendix \ref{mass}). The
 same relations hold for sneutrino sector, with corresponding notations for mixing
 angles
 $s_{\nu_L}, s_{\nu_R}, c_{\nu_L}$ and $c_{\nu_R}$.
 The function $I_3(x,y,z)$ is similar to that mentioned in literature
 \cite{Anna2},
 \bea  I_3(x,y,z)= \frac{xy\log(x/y)+ yz\log(y/z)+
 zx\log(z/x)}{(x-y)(y-z)(z-x)}.\label{I3}\eea

\hs The analytical results appearing in (\ref{deltarho2}) show
that contributions from  two diagrams (e) and (f) to  the
$\Delta^\rho_L$ always  are the same magnitude but opposite in
sign. Therefore the total contribution of  these two diagrams to
$\Delta^\rho_L$ vanishes. On the other hand, results obtained from
the Eq.(\ref{deltarho2}) show that if we neglect the terms of the
slepton mixing, namely $s_L = s_R = s_{\nu_L} =s_{\nu_R} =0 $, the
amounts collected from the $\Delta^\rho_{L,R}$ class diagrams
given in Fig. \ref{rho} are all zero. This corresponds to the case
of lepton flavor conservation: $\Delta^{\rho}_L = \Delta^{\rho}_R
=0.$

\hs We also remind that analytical expressions of other $\Delta$
functions  can be found in Appendix \ref{Diagrams}. These results
demonstrate that the values
 $\Delta^{1\rho}_\mu,
\Delta^{2\rho}_\mu$ and $\Delta^{\rho}_\tau$ given by expressions
(\ref{delta1RhoMu}), (\ref{delta2RhoMu}) and (\ref{deltaRhoTau})
obtained at one loop approximation,
 do not vanish even if
we assume no mixing of the sleptons. These quantities create
non-negligible effects of the lepton masses. They are widely
discussed in many previous papers. Another feature of the SUSYE331
model that we would like to remind here: there are two independent
sources (Yukawa coupling at tree level) to create masses of
slepton and neutrino sectors. Hence, contributions to LFV
corrections come from two independent sources: mixing of lepton
 and  sneutrino sectors. We assume
 that  the model contains  both LFV sources.

Before coming to numerical computation section, it is necessary to
note that the formulas of LFV corrections, such as $\Delta$s in
this case, have not been established for the SUSYE331 model before. So
let us give some general comments on the formulas of $\Delta$s
 which discriminate against those in MSSM versions:
\begin{itemize}
\item  At the one loop approximation, the effective
couplings $\Delta$s  are  obtained from the diagrams such as those
listed in Figs. (\ref{Drhomu1}) and  (\ref{Delrhop}). We can
distinguish two types of diagram which give contribution to
$\Delta$s. The first type of diagram does not include any Higgsino
propagators, for example Fig. \ref{rho} (k) and (l), and they are
known as pure gaugino-mediated diagrams. The second, containing at
least one Higgsino propagator like  remaining diagrams, is
Higgsino-mediated type. In general case, each of these kinds of
diagram  may give main contribution to the $\Delta$s depending on
regions of  mass parameter space.
 If each Higgsino-mediated diagram gives the  dominated contribution to  $\Delta$s that
reach single maximum value. In contrast, each $\Delta$ that gains
values from  pure gaugino-mediated diagrams, $\Delta^{\rho}_{Lk}$
for example, is proportional to $|\mu_\rho|$.
 Additionally,  we can see the analytic expressions of
$\Delta$s given in (\ref{deltarho2}), (\ref{deltarho4}) and
Appendix \ref{Diagrams}. It is well known  in beyond MSSM theories
\cite{Anna1}, all of effective couplings $\Delta$s are obtained
from both types of diagrams, except $\Delta'_{\mu}$. In the limit
of large values of $|\mu_\rho|$, the dominated contributions of
$\Delta$s are caused by pure gaugino-mediated diagrams. This
conclusion also is happened in the SUSYE331. However, in the
SUSYE331, there are  the additional $SU(3)_L$ gaugino-mediated
diagrams. Hence the values of $\Delta^{\rho}_L$ can be changed  in
comparison with other models. Details of this difference are
discussed in section \ref{numerical}.
\item  The difference between the predictions of
the  model under  consideration and other ones due to hypercharge
structure of particle content. For example, let us compare our
expressions of $\Delta$s with those of $\Delta$s in MSMS
\cite{Anna1}. All contributions to the $\Delta$s obtained from
Fig.\ref{rho}, are proportional to $I_3$ functions. Rate
coefficients in both models are the same level for diagrams of
Higgsino-mediated type whereas the rate coefficients in the  model
under  consideration are smaller than that
 in the  MSSM model for diagrams of
  pure gaugino-mediated type.
As a consequence, the large contribution to the $\Delta$s from the
pure gaugino-mediated type will happen if mass parameters are
large. Furthermore, in this limit of mass parameters, the pure
gaugino-mediated diagrams are the only source giving contribution
to radiative corrections $\Delta^{2\rho}_{\mu}$ of muon mass. It
is nature to keep the ratio $Y_{\tau}/Y_{\mu}$, at one loop
correction, to be the same as those
 at tree level. This leads to the limit of the
mass parameters, which  does not exceed $10$ TeV .
\end{itemize}
 In the next section we will investigate some numerical results.
On that basis, we will compare the effects of the LFV origin in
the left- and right-slepton sectors as well as sneutrino sectors.
In order to
 investigate numerically, we are going to use results from \cite{Dong3} such as: $g'/g
 = \frac{3\sqrt{2} s_W}{\sqrt{4 c^2_W-1}} $,
$s^2_W =0.2312$ and $\alpha^{-1}_{em}= 128 $ at the weak scale.

\section{\label{numerical} Numerical results}
\begin{figure}[h]
  \centering
\begin{tabular}{cc}
\epsfig{file=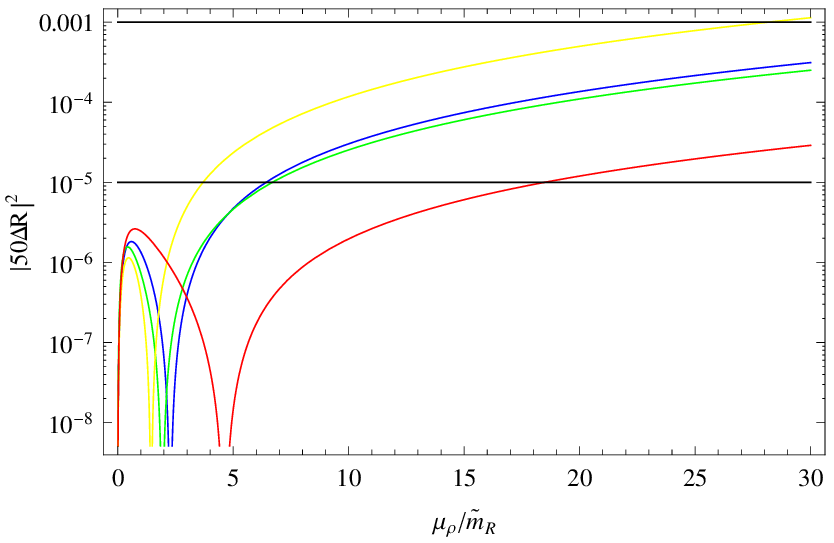,width=0.475\linewidth,clip=} &
\epsfig{file=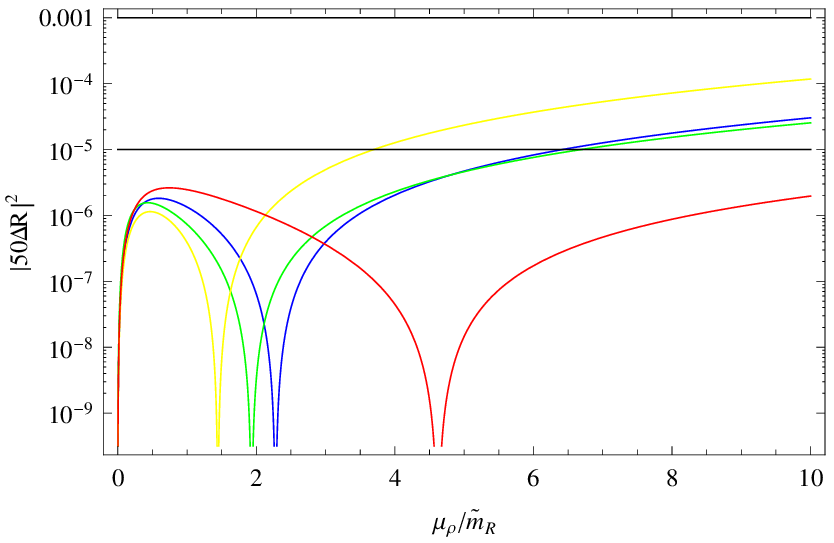,width=0.475\linewidth,clip=} \\
\end{tabular}
  \caption{$|\Delta^{\rho}_R|^2$ as a function of $|\mu_\rho|/\tilde{m}_R$  with
  four different choices of masses ratios: 1) blue curve--$m'=\tilde{m}_R=\tilde{m}_L$ ; 2)
  green curve--$ 3 m'=\tilde{m}_R=\tilde{m}_L$
  ; 3) yellow curve- $m'=\tilde{m}_R= 3 \tilde{m}_L$; 4) red
  curve--$m'=\tilde{m}_R=\tilde{m}_L/3$. Two black lines correspond
  to  two values $10^{-5}$ and $10^{-3}$ of $|50 \Delta^{\rho}_R|^2$
  .}\label{FDeltaRhoR1}
\end{figure}

\begin{figure}[h]
\centering
\begin{tabular}{cc}
\epsfig{file=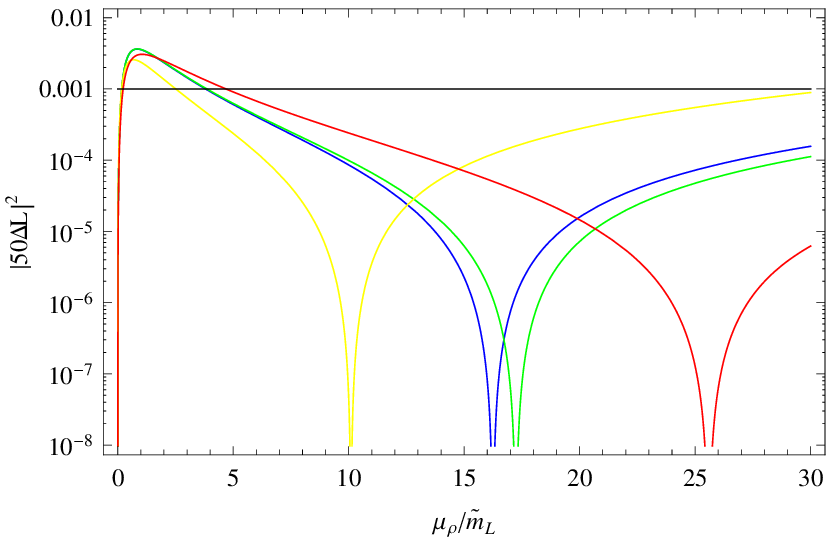,width=0.475\linewidth,clip=} &
\epsfig{file=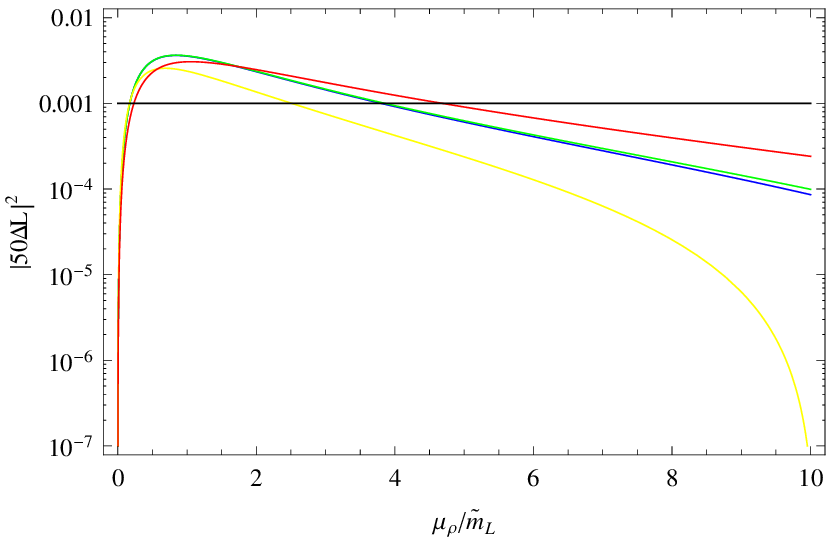,width=0.475\linewidth,clip=} \\
\end{tabular}
\caption{$|\Delta^{\rho}_L|^2$ as a function of
$|\mu_\rho|/\tilde{m}_L$  with
  four different choices of masses ratios: 1) blue curve--$m'=\tilde{m}_R=\tilde{m}_L$ ; 2)
  green curve--$ 3 m'=\tilde{m}_R=\tilde{m}_L$
  ; 3) yellow curve-- $m'=\tilde{m}_L= 3 \tilde{m}_R$; 4) red
  curve--$m'=\tilde{m}_L=\tilde{m}_R/3$. A black line corresponds
  to value $10^{-3}$ of $|50 \Delta^{\rho}_L|^2$.}\label{FDeltaRhoL1}
\end{figure}

In this section we  firstly discuss some numerical results that
relate to any signals of LFV decays $H\rightarrow \mu\tau$. Let us
start with the maximum LFV in both left and right sectors,
especially $s_Lc_L =s_Rc_R =s_{\nu_L}c_{\nu_L} =s_{\nu_R}c_{\nu_R}
=0.5$. It means that we can assign values of mass parameters like:
\bea
     \tilde{m}^2_{(\tau_L, \tau_R,\nu_{\tau_L},
\nu_{\tau_R})} &= &\tilde{m}^2_{(\mu_L,\mu_R ,
\nu_{\mu_L},\nu_{\mu_R})}\crn
 &=&\tilde{m}^2_{L,R,\nu_L, \nu_R}, \crn
 \tilde{m}^2_{(L_2,R_2,
\nu_{L_2},\nu_{R_2}) }&=& 0.2 \, \tilde{m}^2_{L,R,\nu_L, \nu_L}\nn
\eea
 and
 $$\tilde{m}^2_{(L_3,R_3, \nu_{L_3}, \nu_{R_3})}=
1.8~\tilde{m}^2_{(L,R,\nu_L,\nu_R)},$$

where $\tilde{m}^2_{(L,R,\nu_L,\nu_R)}$ are mass parameters used
to compare with SUSY mass scale $m_{SUSY}$. We would like to
emphasize that branching ratios of Higgs decays to muon and tauon
are sizable if $\tan\gamma$ is large enough. Therefore, in the
following calculations, we take $\tan\gamma \sim 50$.

 The Fig.\ref{FDeltaRhoR1} displays the quantity $|50 \Delta^\rho_R|^2$
 as a function of $|\mu_\rho|/\tilde{m}_R$ while
  Fig.\ref{FDeltaRhoL1} displays the $|50 \Delta^\rho_L|^2$
 as a function of $|\mu_\rho|/\tilde{m}_L$
 where  all other relevant parameters are fixed.
 Each curve  presented in Figs.\ref{FDeltaRhoR1},
\ref{FDeltaRhoL1} contains a single peak. All the peaks of the
curves are obtained at mass parameters at which the contribution
of the Higgs-mediated diagrams to $\Delta$s are
dominated. Corresponding to each curve, there are two regions of
mass parameter space separated by deep wells. Deep wells,
which divide the parameter space into two parts. The first part,
the mass parameters are located in the right hand side of deep
wells. In this region of parameter space,the pure gaugino-mediated
type can give main contribution to $\Delta$s. The second part, the
mass parameters are located in the left handed side of deep wells
at which the dominated contribution to $\Delta$s is obtained by
the Higgs-mediated. All of the maximum points of the
curves in the
 Fig.\ref{FDeltaRhoR1} and Fig.\ref{FDeltaRhoL1} are reached at
 $|\mu_\rho|/\tilde{m}_{L,R} \sim \mathcal{O}(10^{-1})$
and these values depend weekly  on the changes of values of
$\tilde{m}_L$ and $\tilde{m}_R$.  On the other hand, the
 maximum values  of the $\Delta^{\rho}_L$ is $\mathcal{O}(10^{-3})$,
 as concerned in the  MSSM
 \cite{Anna1,cruz,Cannoni,lorenzo} while the maximum values of the
 $\Delta^{\rho}_R$ are  much smaller than those  of  $\Delta^{\rho}_L$, specifically
 max$(|\Delta^{\rho}_R|)^2/\mathrm{max} (|\Delta^{\rho}_L|)^2 \sim
10^{-3}$. This large difference comes from the symmetry of
$SU(3)_L \times \mathrm{U}(1)_X$ model. In particular, in the left
side of wells the main contributions to $\Delta^\rho_L$ of
SUSYE331 model come from the $SU(3)_L$ gaugino-mediated diagrams,
namely diagrams ((b), (c), (d)) in Fig.\ref{rho}. In contrast, the
main contributions to $\Delta^\rho_R$ come from only
$\mathrm{U}(1)$ gaugino-mediated diagram.  Figs.\ref{FDeltaRhoR1}
and \ref{FDeltaRhoL1} also show that both $\Delta^{\rho}_{L}$ and
$\Delta^{\rho}_R$ are very sensitive with the changes of
$\tilde{m}_L$ and $\tilde{m}_R$. More details, Fig.\ref{FDeltaRL3}
draws the dependence of $|\Delta^{\rho}_R|^2/|\Delta^{\rho}_L|^2$
on $|\mu_{\rho}|/\tilde{m}_L$, where
$m'=m_\lambda=\tilde{m}_L\equiv m_{SUSY}$ and  four different
fixed values of $\tilde{m}_R$. The
 maximal and minimal values of the ratio
$|\Delta^{\rho}_R|^2/|\Delta^{\rho}_L|^2$ on all the curves in
Fig.\ref{FDeltaRL3} have the same value at different values of
$|\mu_{\rho}|/\tilde{m}_{SUSY}$. In the parameter region where the
Higgs-mediated diagrams give dominated contribution to $\Delta$s,
the ratio $|\Delta^{\rho}_R|^2/|\Delta^{\rho}_L|^2$ is very small
$(< 10^{-3})$. But in the remaining parameters, that ratio is
increased. In the limit $|\mu_{\rho}|/\tilde{m}_{SUSY} \geq 30$,
the ratio $|\Delta^{\rho}_R|^2/|\Delta^{\rho}_L|^2$ reaches a
constant value. More general, we can investigate the influence of
$\tilde{m}_R/\tilde{m}_L$ on the ratio
$|\Delta^{\rho}_R|^2/|\Delta^{\rho}_L|^2$  through contour plots
drawn in Fig.\ref{FDeltaRL4}. On the drawing results showed that
$|\Delta^{\rho}_R|^2/|\Delta^{\rho}_L|^2\leq \mathcal{O}(10^{-2})$
whenever $|\mu_\rho|/m_{SUSY} \leq 5$ and that ratio does not
depend too much on the ratio $\tilde{m}_R /\tilde{ m}_L$. However,
in the limit $|\mu_\rho|/m_{SUSY} \geq 7$ and $\tilde{m}_R
<0.5\tilde{m}_L$,
 the ratio $|\Delta^{\rho}_R|^2/|\Delta^{\rho}_L|^2$
changes very rapidly if small changes $\tilde{m}_L$ and
$\tilde{m}_R$.
It means that chirality effects of phenomena relating with
 $\Delta^{\rho}_L$ and $\Delta^{\rho}_R$  are sensitive with the
 change of ratio  $\tilde{m}_R/\tilde{m}_L$ at large values of
 $\mu_\rho$.
On the other hand, the left picture in Fig.\ref{FDeltaRL4}
indicates that  when  the ratio $|\mu_\rho|/m_{SUSY} \geq 7$, it
will exist in some regions of parameter space of $\tilde{m}_R
,\tilde{m}_L$ at which the contributions of left- and right-lepton
sectors into the $H\rightarrow \mu \tau$ decay process are of the
same order. In this case, the pure gaugino-mediated diagrams give
the dominated contribution to both $\Delta^{\rho}_L$ and
$\Delta^{\rho}_R$, and also $\Delta^{2\rho}_\mu$ ( see
(\ref{delta2RhoMu}) in Appendix \ref{Diagrams}). Recalling that
large values of $\Delta^{2\rho}_\mu$ can strongly affect directly
on the ratio $Y_\mu/Y_\tau$. The results presented in
Fig.\ref{FDeltaRL4} again confirm that whenever
$|\mu_\rho|/m_{SUSY} \geq 7$ and $\tilde{m}_R <0.5\tilde{m}_L$,
the right-lepton sector gives dominated contribution to the
branching ratio of $H \rightarrow \mu \tau$ decay process.

\begin{figure}[h]
\centering
\begin{tabular}{cc}
\epsfig{file=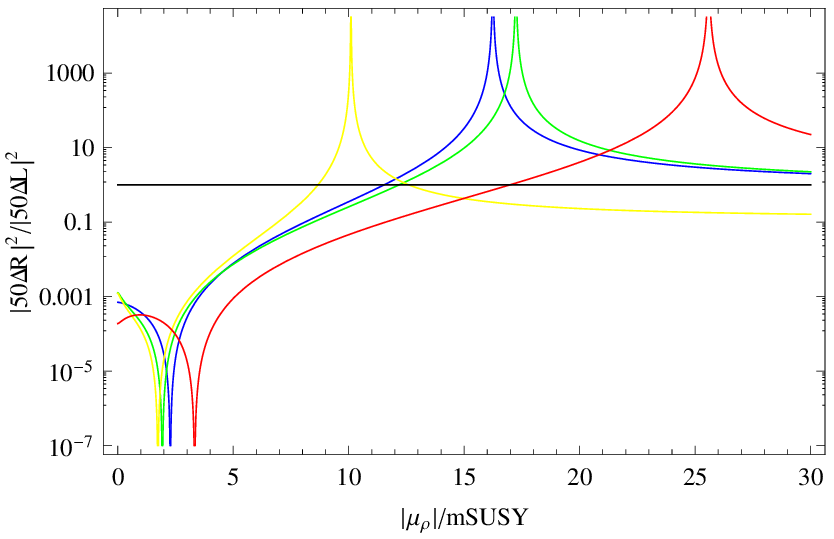,width=0.475\linewidth,clip=} &
\epsfig{file=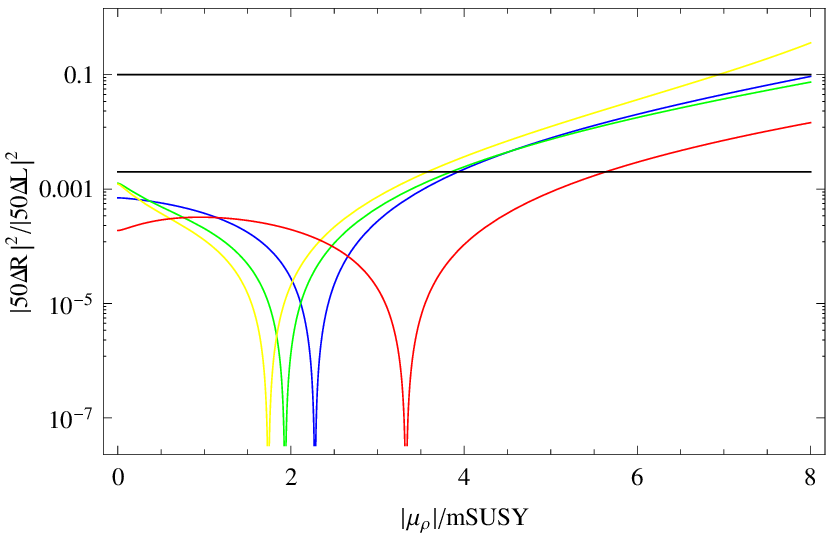,width=0.475\linewidth,clip=} \\
\end{tabular}
\caption{$\frac{|\Delta^{\rho}_R|^2}{| \Delta^{\rho}_L|^2}$ as a
function of $|\mu_\rho|/\tilde{m}_L$ with
  four different choices of masses ratios: 1) blue curve-$m'=\tilde{m}_R=\tilde{m}_L$ ; 2)
  green curve-$ 3 m'=\tilde{m}_R=\tilde{m}_L$
  ; 3) yellow curve- $m'=\tilde{m}_L= 3 \tilde{m}_R$; 4) red
  curve-$m'=\tilde{m}_L=\tilde{m}_R/3$.
  A black line in the left side of figure corresponding
to  the value  $\frac{|\Delta^{\rho}_R|^2}{|
  \Delta^{\rho}_L|^2}$ equals 1. Both black lines in the right side
 of figure presenting $\frac{|\Delta^{\rho}_R|^2}{| \Delta^{\rho}_L|^2}$
 are  $2 \times10^{-3}$ and $0.1$.}
  \label{FDeltaRL3}
\end{figure}
\begin{figure}[h]
\centering
\begin{tabular}{cc}
\epsfig{file=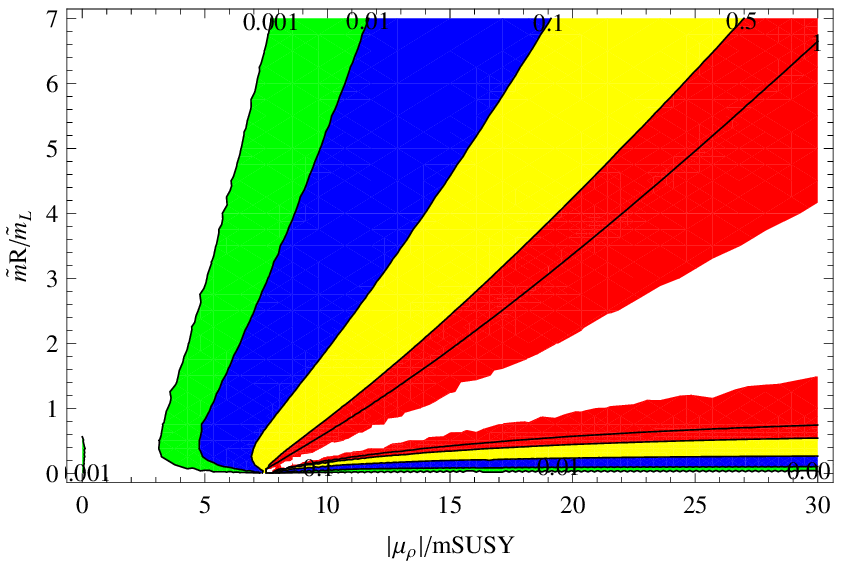,width=0.47\linewidth,clip=} &
\epsfig{file=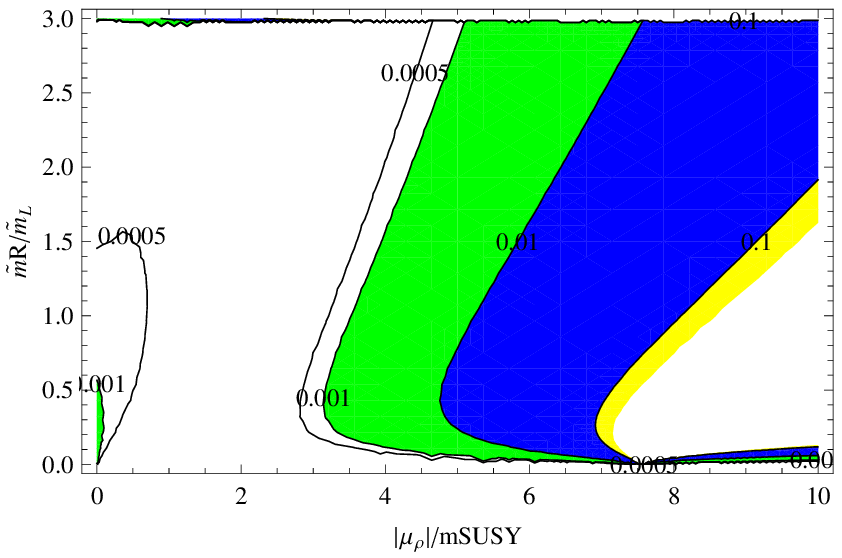,width=0.47\linewidth,clip=} \\
\end{tabular}
\caption{ Contour plot of $\frac{|\Delta^{\rho}_R|^2}{|
\Delta^{\rho}_L|^2}$, $\tilde{m}_R/\tilde{m}_L$  vs
$|\mu_\rho|/m_{SUSY}$, where $\tilde{m}_R= \tilde{m}_{\nu_R}$,
$m'=m_{\lambda}=\tilde{m}_L=\tilde{m}_{\nu_L}=m_{SUSY}$. The red
region corresponds to the values of $\frac{|\Delta^{\rho}_R|^2}{|
\Delta^{\rho}_L|^2} \geq 0.5$.
  }\label{FDeltaRL4}
\end{figure}

\begin{figure}[h]
\centering
\begin{tabular}{cc}
\epsfig{file=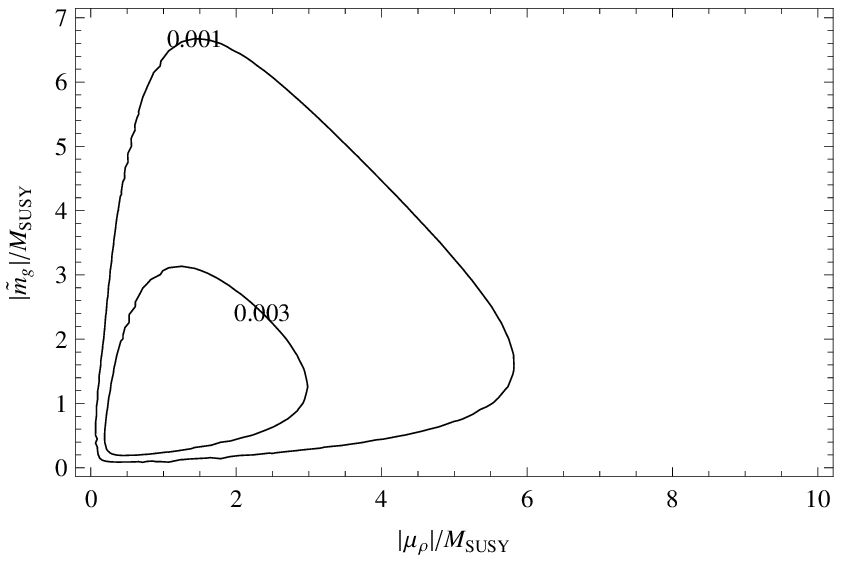,width=0.47\linewidth,clip=} &
\epsfig{file=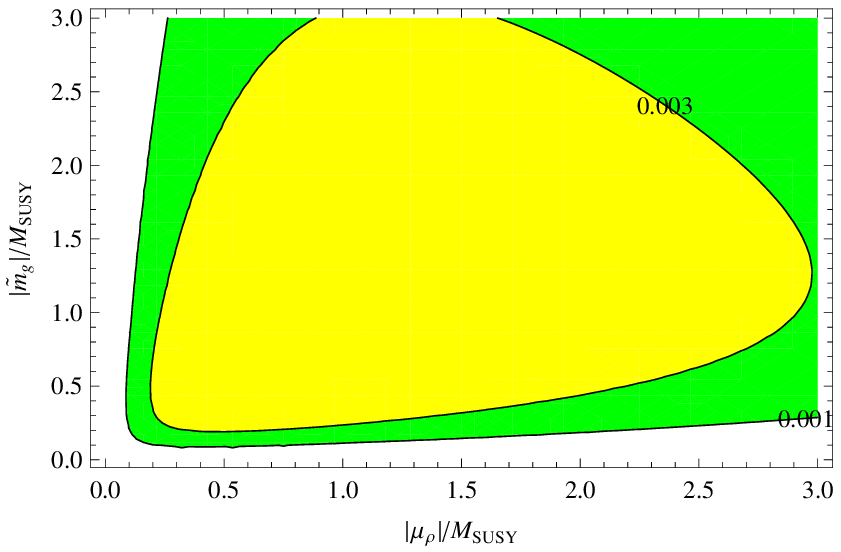,width=0.47\linewidth,clip=} \\
\end{tabular}
\caption{ Contour plot of $BR(H\rightarrow
\mu\tau)/BR(H\rightarrow \tau\tau)$, $\tilde{m}_g$ vs
$|\mu_\rho|/m_{SUSY}$, where $m'=m_{\lambda}=\tilde{m}_g$ and
$\tilde{m}_R=\tilde{m}_{\nu_R}=\tilde{m}_L=\tilde{m}_{\nu_L}=m_{SUSY}$.
In the left picture, the green and yellow regions correspond  to
the values of $BR(H\rightarrow \mu\tau)/BR(H\rightarrow \tau\tau)
\geq \mathcal{O}(10^{-3})$ .
  }\label{FDeltaRL5}
\end{figure}


\hs Now we investigate more details  the region of parameter space
where Higgsino-mediated diagrams give a dominated contribution. In
this region of parameter space, both $\Delta^{2\rho}_{\mu}$ and
$\Delta^{\rho}_R$ are much smaller than $\Delta^{\rho}_L$ so we
just focus on $\Delta^{\rho}_L$.  From Fig.\ref{FDeltaRL5}, we can
estimate the ratio of
 $Br(H\rightarrow \mu\tau)/Br(H\rightarrow  \tau\tau)$ that
 can reach the order of $10^{-3}$ in the limit  $0.1 \leq |\mu_{\rho}|/M_{SUSY}
  \leq 6$ and $0.1 \leq |\tilde{m}_{g}|/M_{SUSY}
  \leq 7$ where  $\tilde{m}_{g}$ is mass of gauginos.
Let us  briefly review the decay properties of neutral Higgs
bosons in the SUSYE331 model. At the tree level, the couplings of
  neutral Higgs bosons to up-fermions, down-fermion
  are modified with respect to the SM  coupling by factors which
  are given in Table \ref{table1}.
   \begin{table}[h]
\caption{ Coupling of neutral Higgs bosons to
fermion.}\label{table1}
  \begin{tabular}{|c |c|c|c|c|}
    \hline
   Particles & Up-fermion& Down-fermion & Exotic up-quark &
   Exotic down-quark\\
    \hline
    SM  Higgs & 1 & 1 & 0 &0\\
    $\varphi_{Sa36}$ & $c_{\alpha}$ & $c_{\alpha}$
     &$ s_{\alpha}/s_{\gamma}$&$ c_{\alpha}/s_{\gamma}$ \\
    $\phi_{Sa36}$ & $s_{\alpha} $& $s_{\alpha} $& $c_{\alpha}$&
    $s_{\alpha}$\\
    \hline
     \end{tabular}
\end{table}

  We assume that  all exotic quarks have  masses  heavier than
  that of all neutral Higgses. It means that the neutral Higgs cannot
  decay into the exotic quarks. The neutral Higgs bosons
  may  decay mainly into the pairs of fermions.
 This prediction depends on the mass of the neutral Higgs.
 For neutral Higgs $\varphi_{Sa36}$, its mass depends on the
 vacuum expectation values $v, v^\prime$. So it should be
 predicted SM Higgs with mass
   smaller than about $130$ GeV.
 Decay of $\varphi_{Sa36}$ to $b\overline{b}$
 and $\tau \overline{\tau}$ are dominated, the branching ratios
 of $90 $ percent and $8$ percent, respectively.
Combined with the results in  Fig.\ref{FDeltaRL5}, the branching
ratio $Br(\varphi_{Sa36}\rightarrow \mu\tau)$ is $8 \times
10^{-3}$ percent. This  may be a good signification of new physics
in the present limits of colliders. For neutral Higgs
$\phi_{Sa36}$, it is heavy Higgs, the main productions of decay
are the the gauge bosons such as $W^+ W^-$, $Z Z$,...Hence, the
branching ratio $\phi \rightarrow \mu \tau$ is very suppressed.
 We would like to note that the effective interactions
  of the muon, tauon and Higgs given in (\ref{FCNC3})
 not only leads to the LFV of Higgs decay process,
 but also affects the other physical processes
 with lepton-flavor violations.
 Some of these processes which are looked seriously
  by present experiments
 are, for instance,  $\tau \rightarrow \mu\mu\mu$
 and $\tau \rightarrow \mu\gamma$. Let us  apply
  the effective couplings given in (\ref{FCNC3}) to the $\tau\rightarrow
 \mu \mu \mu$ decay process.
 In a general way-regardless of the model, the general
 effective Lagrangian describing decay of $\tau
\rightarrow \mu \mu \mu$ was studied in  \cite{okada1}.  However,
in this work we focus on the effect of the Higgs-mediated LFV
interactions on the $\tau \rightarrow \mu \mu \mu$ decay process.
Hence, the four dimensional effective Lagrangian which is built
through Higgs exchange is formulated by \bea
\mathcal{L}^{eff}_{\tau\mu\mu\mu}&=& -2 \sqrt{2}G_F m_\mu m_\tau
\tan\gamma \left(\frac{s^2_\alpha}{m^2_{\phi_{Sa36}}}
+\frac{c^2_\alpha}{m^2_{\varphi_{Sa36}}}\right) (\mu^c\mu+
\bar{\mu} \bar{\mu}^c)\crn &\times& (\Delta^{\rho}_L \tau^c\mu+
\Delta^{\rho}_R \mu^c\tau ) + \mathrm{H.c.}. \label{tau3mu1}\eea
We would like to remind that the decay process of $\tau
\rightarrow \mu \mu \mu$ were investigated, by \cite{Anna2,okada1}
for examples, in a general model-independent way.
 The predicted results  show that when
   Higgs exchange effects are much smaller than
  other ones, the ratio $Br(\tau\rightarrow 3\mu)/BR(\tau\rightarrow \mu\gamma)$
becomes constant with a value $\sim \mathcal{O}(10^{-3})$. Now, we
will discuss in  more details whether Higgs-mediated effects can
make any significations to the ratio $Br(\tau\rightarrow
3\mu)/BR(\tau\rightarrow \mu\gamma)$ in the SUSYE331 model.
 We can divide our results into two cases, namely
 $\phi^*_{S_{a36}}$  and $\varphi^*_{S_{a36}}$
 Higgs-mediated effects. The results can be written in two respective forms:
\bea BR(\tau^-\rightarrow\mu^-\mu^+\mu^-)_{\phi^*_{S_{a36}}}
&=&\frac{1}{8} \tan^2\gamma~
\frac{m^2_{\mu}m^2_{\tau}}{m^4_{\phi_{S_{a36}}}} s^4_{\alpha}
\left( |\Delta^\rho_L|^2+ |\Delta^\rho_R|^2\right) \crn &\simeq&
7\times 10^{-11}  \left(\frac{\tan\gamma}{40}\right)^2 \left(
\frac{100\mathrm{GeV}}{m_{\phi_{S_{a36}}}}\right)^4\crn &\times&
\left(\frac{|\Delta^\rho_L|^2+ |\Delta^\rho_R|^2}{10^{-3}} \right)
s^4_\alpha\crn \label{tau3muhiggs1}\eea
 and
 \bea BR(\tau^-\rightarrow\mu^-\mu^+\mu^-)_{\varphi^*_{S_{a36}}} &=& \frac{1}{8}
\tan^2\gamma~  \frac{m^2_{\mu}m^2_{\tau}}{m^4_{\varphi_{S_{a36}}}}
c^4_{\alpha} \left( |\Delta^\rho_L|^2+ |\Delta^\rho_R|^2\right)
\crn &\simeq& 7\times 10^{-11}
\left(\frac{\tan\gamma}{40}\right)^2 \left(
\frac{100\mathrm{GeV}}{m_{\varphi_{S_{a36}}}}\right)^4\crn
&\times& \left(\frac{|\Delta^\rho_L|^2+
|\Delta^\rho_R|^2}{10^{-3}} \right) c^4_\alpha.
\label{tau3muhiggs2}\eea
 These results  immediately  lead to a consequence: the maximum
 contribution of Higgs exchange processes can be estimated
 through the formula:
\bea BR(\tau^-\rightarrow\mu^-\mu^+\mu^-)_{H^*} &\simeq& 7\times
10^{-11}  \left(\frac{\tan\gamma}{40}\right)^2 \left(
\frac{100\mathrm{GeV}}{m_{\mathrm{H}}}\right)^4\crn &\times&
\left(\frac{|\Delta^\rho_L|^2+ |\Delta^\rho_R|^2}{10^{-3}}
\right). \label{tau3muhiggs3}\eea
 The values of the branching ratios decrease rapidly corresponding to the
 enhancement of Higgs masses . We stress that in the
 model under  consideration, the $\varphi_{S_{a36}}$ is identified with the
SM  Higgs boson and the remain, $\phi_{S_{a36}}$, is heavy one.
Overall, in our model, the SM  Higgs-mediated gives larger
contribution to the branching ratio of the $\tau \rightarrow \mu
\mu \mu$ decay process than that of heavy Higgs. That kind of
branching increases if the $\tan \gamma$ increases. The branching
ratio estimated in (\ref{tau3muhiggs3}) is $ \simeq 10^{-11}$  in
the limit of $\tan\gamma \simeq 50$ and the Higgs mass is of the
order of $100$ GeV. However, the branching ratio of $\tau
\rightarrow\mu\mu\mu$ can be reached at the present limits of
experiment  $BR(\tau^- \rightarrow \mu^-\mu^+\mu^-) \leq 3.2
\times 10^{-8}$ \cite{belle1}. It means that the contribution to
$\tau \rightarrow\mu\mu\mu$ is suppressed in the limit of
$\tan\gamma \simeq 50$. This result is different from that
predicted in the MSSM model \cite{Babu,Anna1,Anna2}. In
particular, for the MSSM, the dominant contributions to $BR(\tau
\rightarrow \mu \mu \mu)$ are induced by the dipole term  and the
Higgs-mediated term at the limit $\tan \beta = 50, m_A= 100$ GeV.

Because of sub-dominated contribution of Higgs-mediated to $BR(
\tau \rightarrow\mu\mu\mu )$, the dominated contribution to that
branching ratio is still obtained from the photon-penguin
couplings. This result leads to the values of well-known ratios
such as
\bea \frac{Br(\tau\rightarrow 3\mu)}
 {BR(\tau\rightarrow \mu\gamma)}\simeq \mathcal{O}(10^{-3})
 \label{huong2}\eea
 The predicted result is  the concerned result given in \cite{Babu,Anna2,okada1,hisano}.

 We emphasize that in the SUSYE331 model, in order to get the
 dominated contribution to the  $B(\tau \rightarrow \mu \mu \mu)$,
 the values of $\tan \gamma$ must be $10^{2}$. In this limit of $\tan \gamma$, the
 result given in (\ref{huong2}) is not holden.

\section{\label{concl}Conclusions}
In this paper, we   have studied the LFV interactions of Higgs
bosons in the SUSYE331 model.  We have   the unique existence of
the lepton-number violation  in the slepton sector at the tree
level. On the basis of this assumption we have examined the
lepton-number violating  interactions of Higgs bosons
 at the  one-loop  level.
Specially we have  concentrated our  study  on the LFV couplings
of Higgs bosons with  muon and  tauon. The analytical expressions
of the effective Higgs-muon-tauon couplings are established at the
one-loop level. One of the features  is that the  model does not
 contain the LFV interactions of neutral pseudo-scalar Higgs bosons.
For the neutral Higgs scalars, the model contains two types of
radiative interactions that violate lepton number, namely,
$\phi_{Sa36} \mu \tau$  and $\varphi_{Sa36} \mu \tau$. These
effective couplings depend on  ratios of SUSY mass parameters and
$\tan \gamma$. There is an exactly  similar to the other SUSY
models, all LFV couplings are built from two types of diagrams as
Higgs-mediated diagrams and pure gaugino-mediated ones. Depending
on the SUSY parameters, each type of diagram gives the main
contribution to LFV couplings. In this work, we have also studied
the branching ratio of the neutral Higgs decay into muon and
tauon. In the limit $|\mu_\rho|/m_{SUSY} \leq 7$, the ratio of
$BR(H \rightarrow \tau\mu)/BR(H \rightarrow \tau\tau)$ in this
region can reach values that can be observed by near future
experiments and the contributions from both left and right LFV
sectors to $Br(H \rightarrow \mu \tau) $ are of the same order.
Outside this region the effects of left and right LFV terms mix in
different ways in different regions of mass parameter space. We
predicted that for the SM Higgs boson, LHC may detect the decay of
SM  Higgs boson to muon and tauon. For heavy Higgs bosons, the
branching of LFV decay is very suppressed. We have also studied
the contribution of Higgs exchange to decay $H\rightarrow 3\mu$.
In the limit $\tan \gamma = 50$,  the $Br(H\rightarrow 3 \mu)$  is
very small, out of direct detection of present searching of
experiment and it leads to predicted results such as the ratio of
$BR(\tau\rightarrow3\mu)/BR(\tau\rightarrow \mu\gamma)$.

 \section*{Acknowledgments}
 L.T.H would like to thank the Organizers of KEK-Vietnam Visiting
 Program 2011 (Exchange Program for East Asia Young Researchers,
 JSPS), especially Prof. Y. Kurihara,  for the support for his
 initial work at KEK. This work was supported in part by the National Foundation for
Science and Technology Development (NAFOSTED) of Vietnam under
Grant No. 103.01-2011.63.
\\[0.3cm]

\appendix
\section{\label{Diagrams} Analytic formulas and diagrams
 {\bf contributing}  to $\Delta$s}

 Let us write down all expressions of $\Delta$s given as following
\bea \Delta^{1\rho}_{\mu}
 &=&-\frac{g^{\prime 2}}{108 \pi^2}
\mu_{\rho} m' \left[ c^2_L  I_3(m^{\prime 2}, \mu^2_{\rho},
\tilde{m}^2_{L_2})+ s^2_L I_3 (m^{\prime 2}, \mu^2_{\rho},
\tilde{m}^2_{L_3}) \right]\crn
&-& \frac{ g^{2}}{24 \pi^2}\mu_{\rho} m_{\lambda} \left[ c^2_L
I_3(m^{2}_{\lambda}, \mu^2_{\rho}, \tilde{m}^2_{L_2})+ s^2_LI_3
(m^{ 2}_{\lambda}, \mu^2_{\rho}, \tilde{m}^2_{L_3}) \right]\crn
&-&\frac{g^{2}}{16 \pi^2} \mu_{\rho} m_{\lambda} \left[
c^2_{\nu_L} I_3(m^{2}_{\lambda}, \mu^2_{\rho},
\tilde{m}^2_{\nu_{L2}})+ s^2_{\nu_L} I_3 (m^{2}_{\lambda},
\mu^2_{\rho}, \tilde{m}^2_{\nu_{L3}}) \right]\crn
&-&\frac{g^{2}}{16 \pi^2} \mu_{\rho} m_{\lambda} \left[
c^2_{\nu_R} I_3(m^{2}_{\lambda}, \mu^2_{\rho},
\tilde{m}^2_{\nu_{R2}})+ s^2_{\nu_R} I_3 (m^{2}_{\lambda},
\mu^2_{\rho}, \tilde{m}^2_{\nu_{R3}}) \right]\crn
&+&
  \frac{Y_{\nu_{\mu\tau}}(h_{\mu\tau}-h_{\tau\mu})
  \mu_{\rho}}{8 \pi^2}\crn
 &\times& \left[s^2_{\nu_{(L-R)}}
 \left( I_3(\mu^2_{\rho},\tilde{m}^2_{\nu_{L2}},
 \tilde{m}^2_{\nu_{R2}})+ I_3(\mu^2_{\rho},\tilde{m}^2_{\nu_{L3}},
 \tilde{m}^2_{\nu_{R3}})  \right)\right.\crn
&+&  \left. c^2_{\nu_{(L-R)}}
\left(I_3(\mu^2_{\rho},\tilde{m}^2_{\nu_{L3}},
 \tilde{m}^2_{\nu_{R2}})+ I_3(\mu^2_{\rho},\tilde{m}^2_{\nu_{L2}},
 \tilde{m}^2_{\nu_{R3}})  \right) \right]\crn
 &+& \frac{g^{\prime 2}}{288\pi^2} \mu_{\rho}
m' \left[c^2_L \left( c^2_R I_3(m^{\prime 2},
\tilde{m}^2_{L_2},\tilde{m}^2_{R_2})+s^2_R I_3(m^{\prime 2},
\tilde{m}^2_{L_2},\tilde{m}^2_{R_3} ) \right)\right.\crn
&+& \left. s^2_L \left(c^2_R I_3(m^{\prime 2},
\tilde{m}^2_{L_3},\tilde{m}^2_{R_2})+ s^2_RI_3(m^{\prime 2},
\tilde{m}^2_{L_3},\tilde{m}^2_{R_3} )\right) \right].
  \label{delta1RhoMu} \eea
\bea \Delta^{2\rho}_{\mu}
 &=&  \frac{g^{\prime 2}}{288\pi^2} \mu_{\rho}
m' s_Lc_Ls_Rc_R\left[ I_3(m^{\prime 2},
\tilde{m}^2_{L_2},\tilde{m}^2_{R_2})- I_3(m^{\prime 2},
\tilde{m}^2_{L_2},\tilde{m}^2_{R_3}) \right.\crn
&-& \left.  I_3(m^{\prime 2},
\tilde{m}^2_{L_3},\tilde{m}^2_{R_2})+ I_3(m^{\prime 2},
\tilde{m}^2_{L_3},\tilde{m}^2_{R_3} ) \right].
\label{delta2RhoMu}\eea
\bea \Delta^{\rho}_{\tau}
 &=&
 -\frac{g^{\prime 2}}{108 \pi^2}
\mu_{\rho} m' \left[ s^2_L  I_3(m^{\prime 2}, \mu^2_{\rho},
\tilde{m}^2_{L_2})+ c^2_L I_3 (m^{\prime 2}, \mu^2_{\rho},
\tilde{m}^2_{L_3}) \right]\crn
&-& \frac{g^{2}}{24\pi^2}\mu_{\rho} m_{\lambda} \left[ s^2_L
I_3(m^{2}_{\lambda}, \mu^2_{\rho}, \tilde{m}^2_{L_2})+ c^2_LI_3
(m^{ 2}_{\lambda}, \mu^2_{\rho}, \tilde{m}^2_{L_3}) \right]\crn
&-&\frac{g^{2}}{16 \pi^2} \mu_{\rho} m_{\lambda} \left[
s^2_{\nu_L} I_3(m^{2}_{\lambda}, \mu^2_{\rho},
\tilde{m}^2_{\nu_{L2}})+ c^2_{\nu_L} I_3 (m^{2}_{\lambda},
\mu^2_{\rho}, \tilde{m}^2_{\nu_{L3}}) \right]\crn
&-&\frac{g^{2}}{16 \pi^2} \mu_{\rho} m_{\lambda} \left[
s^2_{\nu_R} I_3(m^{2}_{\lambda}, \mu^2_{\rho},
\tilde{m}^2_{\nu_{R2}})+ c^2_{\nu_R} I_3 (m^{2}_{\lambda}
\mu^2_{\rho}, \tilde{m}^2_{\nu_{R3}}) \right]\crn
&+&
  \frac{Y_{\nu_{\mu\tau}}(h_{\mu\tau}-h_{\tau\mu})
  \mu_{\rho}}{8 \pi^2}\crn
 &\times& \left[s^2_{\nu_{(L-R)}}
  \left(I_3(\mu^2_{\rho},\tilde{m}^2_{\nu_{L2}},
 \tilde{m}^2_{\nu_{R2}})+I_3(\mu^2_{\rho},\tilde{m}^2_{\nu_{L3}},
 \tilde{m}^2_{\nu_{R3}})  \right)\right.\crn
&+&  \left. c^2_{\nu_{(L-R)}}
\left(I_3(\mu^2_{\rho},\tilde{m}^2_{\nu_{3L}},
 \tilde{m}^2_{\nu_{R2}})+ I_3(\mu^2_{\rho},\tilde{m}^2_{\nu_{L2}},
 \tilde{m}^2_{\nu_{R3}})  \right) \right]\crn
 &+& \frac{g^{\prime 2}}{288\pi^2} \mu_{\rho}
m' \left[s^2_L \left( s^2_R I_3(m^{\prime 2},
\tilde{m}^2_{L_2},\tilde{m}^2_{R_2})+c^2_R I_3(m^{\prime 2},
\tilde{m}^2_{L_2},\tilde{m}^2_{R_3} ) \right)\right.\crn
&+& \left. c^2_L \left(s^2_R I_3(m^{\prime 2},
\tilde{m}^2_{L_3},\tilde{m}^2_{R_2})+ c^2_RI_3(m^{\prime 2},
\tilde{m}^2_{L_3},\tilde{m}^2_{R_3} )\right)
\right].\label{deltaRhoTau} \eea


\bea \Delta^{1\rho'}_\mu&=&
\frac{Y^2_{\nu_{\mu\tau}}}{4\pi^2}\mu^2_{\rho} \left[
s^2_{\nu_{(L-R)}} \left( I_3(\mu^2_\rho,
\tilde{m}^2_{\nu_{L2}},\tilde{m}^2_{\nu_{R2}})+I_3(\mu^2_\rho,
\tilde{m}^2_{\nu_{L3}},\tilde{m}^2_{\nu_{R3}})\right)\right.\crn
&+& \left.c^2_{\nu_{(L-R)}} \left( I_3(\mu^2_\rho,
\tilde{m}^2_{\nu_{L2}},\tilde{m}^2_{\nu_{R3}})+I_3(\mu^2_\rho,
\tilde{m}^2_{\nu_{L3}},\tilde{m}^2_{\nu_{R2}})\right) \right].
\label{delta1rhop}\eea
 \bea \Delta^{2\rho'}_{\mu} &=& \frac{g^{\prime 2} m'}{144
 \pi^2}\crn
 &\times&\left[ \frac{h'_\mu c_L c_R + h'_\tau s_L s_R+ h'_{\mu\tau}
  c_L s_R+  h'_{\tau\mu} s_L
 c_R }{Y_\tau} c_L c_R I_3(m^{\prime 2},
 \tilde{m}^2_{L_2},\tilde{m}^2_{R_2}) \right.\crn
 &+&\left.\frac{h'_\mu s_L s_R + h'_\tau c_L c_R- h'_{\mu\tau}
 s_L c_R -h'_{\tau\mu} c_L
 s_R }{Y_\tau} s_L s_R I_3(m^{\prime 2},
 \tilde{m}^2_{L_3}, \tilde{m}^2_{R_3}) \right.\crn
 &-&\left.\frac{-h'_\mu s_L c_R + h'_\tau c_L s_R- h'_{\mu\tau}
 s_L s_R +h'_{\tau\mu} c_L
 c_R }{Y_\tau} s_L c_R I_3(m^{\prime 2},
 \tilde{m}^2_{L_3}, \tilde{m}^2_{R_2}) \right.\crn
 &-&\left. \frac{-h'_\mu c_L s_R + h'_\tau s_L c_R+ h'_{\mu\tau}
 c_L c_R -h'_{\tau\mu} s_L
 s_R }{Y_\tau} c_L s_R I_3(m^{\prime 2},
 \tilde{m}^2_{L_2}, \tilde{m}^2_{R_3})
  \right]. \crn\label{deltaPrRhoPrMu}
 \eea
 \bea \Delta^{\rho'}_{\tau} &=&  \Delta^{1\rho'}_{\mu} + \frac{g^{\prime 2} m'}{144
 \pi^2}\crn
 &\times&\left[ \frac{h'_\mu c_L c_R + h'_\tau s_L s_R
 + h'_{\mu\tau} c_L s_R+  h'_{\tau\mu} s_L
 c_R }{Y_\tau} s_L s_R I_3(m^{\prime 2},
 \tilde{m}^2_{L_2},\tilde{m}^2_{R_2}) \right.\crn
 &+&\left.\frac{h'_\mu s_L s_R + h'_\tau c_L c_R
 - h'_{\mu\tau} s_L c_R -h'_{\tau\mu} c_L
 s_R }{Y_\tau} c_L c_R I_3(m^{\prime 2},
 \tilde{m}^2_{L_3}, \tilde{m}^2_{R_3}) \right.\crn
 &+&\left.\frac{-h'_\mu s_L c_R + h'_\tau c_L s_R- h'_{\mu\tau}
  s_L s_R +h'_{\tau\mu} c_L
 c_R }{Y_\tau} c_L s_R I_3(m^{\prime 2},
 \tilde{m}^2_{L_3}, \tilde{m}^2_{R_2}) \right.\crn
 &+&\left. \frac{-h'_\mu c_L s_R + h'_\tau s_L c_R+ h'_{\mu\tau}
  c_L c_R -h'_{\tau\mu} s_L
 s_R }{Y_\tau} s_L c_R I_3(m^{\prime 2},
 \tilde{m}^2_{L_2}, \tilde{m}^2_{R_3})
  \right] \crn \label{deltaRhoPrTau} \eea
In the case of $\Delta^{\rho'}_L$, it also receives contributions
from two diagrams (similar to diagrams  (e) and (f) in
Fig.\ref{rho} ) which cancel each other. Therefore we do not
repeat them in Figs.and \ref{Delrhop1} and \ref{rhop} . Formula of
$\Delta^{\rho'}_L$ then is
%
 \bea \Delta^{\rho'}_{L} &=& \frac{g^{\prime 2} m'}{144
 \pi^2}\crn
 &\times&\left[ \frac{h'_\mu c_L c_R + h'_\tau s_L s_R
 + h'_{\mu\tau} c_L s_R+  h'_{\tau\mu} s_L
 c_R }{Y_\tau} c_L s_R I_3(m^{\prime 2},
 \tilde{m}^2_{L_2},\tilde{m}^2_{R_2}) \right.\crn
 &-&\left.\frac{h'_\mu s_L s_R + h'_\tau c_L c_R
 - h'_{\mu\tau} s_L c_R -h'_{\tau\mu} c_L
 s_R }{Y_\tau} s_L c_R I_3(m^{\prime 2},
 \tilde{m}^2_{L_3}, \tilde{m}^2_{R_3}) \right.\crn
 &-&\left.\frac{-h'_\mu s_L c_R + h'_\tau c_L s_R
 - h'_{\mu\tau} s_L s_R +h'_{\tau\mu} c_L
 c_R }{Y_\tau} s_L s_R I_3(m^{\prime 2},
 \tilde{m}^2_{L_3}, \tilde{m}^2_{R_2}) \right.\crn
 &+&\left. \frac{-h'_\mu c_L s_R + h'_\tau s_L c_R
 + h'_{\mu\tau} c_L c_R -h'_{\tau\mu} s_L
 s_R }{Y_\tau} c_L c_R I_3(m^{\prime 2},
 \tilde{m}^2_{\tilde{L}_2}, \tilde{m}^2_{\tilde{R}_3})
  \right]. \crn \label{deltaRhoPrL} \eea
 \bea \Delta^{\rho'}_{R} &=&\frac{g^{\prime 2} m'}{144
 \pi^2}\crn
 &\times&\left[ \frac{h'_\mu c_L c_R + h'_\tau s_L s_R
 + h'_{\mu\tau} c_L s_R+  h'_{\tau\mu} s_L
 c_R }{Y_\tau} s_L c_R I_3(m^{\prime 2},
 \tilde{m}^2_{L_2},\tilde{m}^2_{R_2}) \right.\crn
 &-&\left.\frac{h'_\mu s_L s_R + h'_\tau c_L c_R
 - h'_{\mu\tau} s_L c_R -h'_{\tau\mu} c_L
 s_R }{Y_\tau} c_L s_R I_3(m^{\prime 2},
 \tilde{m}^2_{L_3}, \tilde{m}^2_{R_3}) \right.\crn
 &+&\left.\frac{-h'_\mu s_L c_R + h'_\tau c_L s_R
 - h'_{\mu\tau} s_L s_R +h'_{\tau\mu} c_L
 c_R }{Y_\tau} c_L c_R I_3(m^{\prime 2},
 \tilde{m}^2_{L_3}, \tilde{m}^2_{R_2}) \right.\crn
 &-&\left. \frac{-h'_\mu c_L s_R + h'_\tau s_L c_R
 + h'_{\mu\tau} c_L c_R -h'_{\tau\mu} s_L
 s_R }{Y_\tau} s_L s_R I_3(m^{\prime 2},
 \tilde{m}^2_{L_2}, \tilde{m}^2_{R_3})
  \right]. \crn \label{deltaRhoPrR} \eea


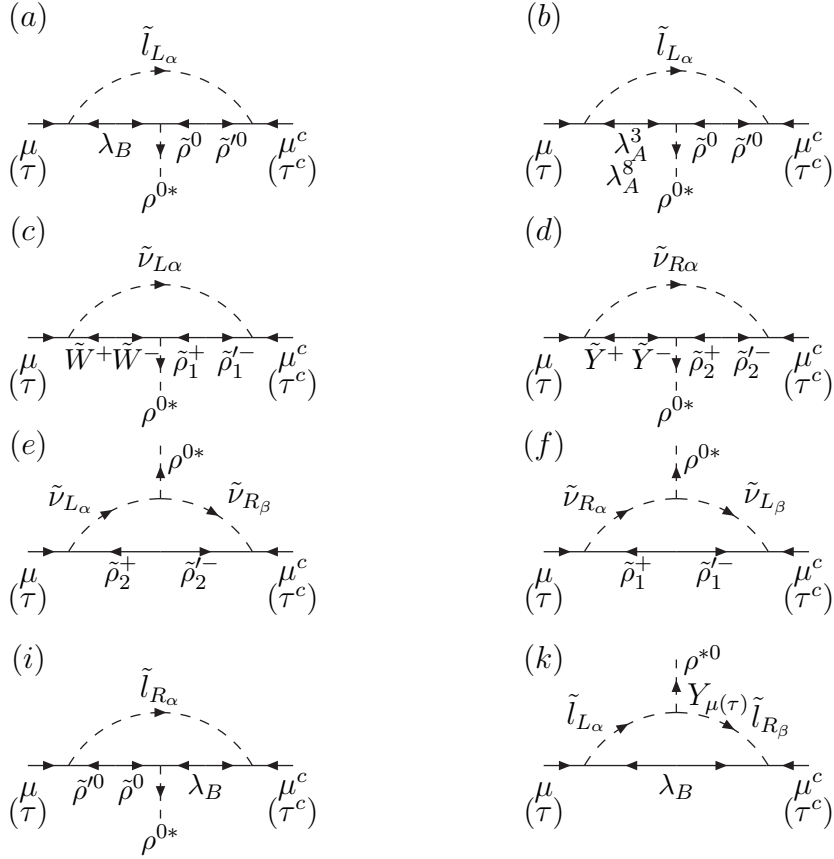
\begin{figure}[t]
\vspace{0.4cm}
\begin{center}
\begin{picture}(120,80)(-60,-40)
\ArrowLine(-50,0)(-34,0) \ArrowLine(-17,0)(-34,0)
\ArrowLine(-17,0)(0,0) \ArrowLine(17,0)(0,0)
\ArrowLine(17,0)(34,0) \ArrowLine(50,0)(34,0)
\DashArrowArcn(0,-20)(40,150,30){4} \DashArrowLine(0,0)(0,-20){3}
\Text(-50,-10)[]{$\mu$} \Text(50,-8)[]{$\mu^c$}
\Text(-50,-18)[]{$(\tau)$} \Text(50,-18)[]{$(\tau^c)$}
\Text(27,-9)[]{\small $\tilde{\rho}^{\prime0}$}
\Text(11,-9)[]{\small $\tilde{\rho}^{0}$}
\Text(-17,-8)[]{\small$\lambda_B$}
 \Text(0,-28)[]{\small
$\rho^{0*}$} \Text(0,30)[]{$\tilde{l}_{L_\alpha}$}
\Text(-50,40)[]{$(a)$}
\end{picture}
\hglue 2.5cm
\begin{picture}(120,80)(-60,-40)
\ArrowLine(-50,0)(-34,0) \ArrowLine(-17,0)(-34,0)
\ArrowLine(-17,0)(0,0) \ArrowLine(17,0)(0,0)
\ArrowLine(17,0)(34,0) \ArrowLine(50,0)(34,0)
\DashArrowArcn(0,-20)(40,150,30){4} \DashArrowLine(0,0)(0,-20){3}
\Text(-50,-10)[]{$\mu$} \Text(50,-8)[]{$\mu^c$}
\Text(-50,-18)[]{$(\tau)$} \Text(50,-18)[]{$(\tau^c)$}
\Text(27,-9)[]{\small $\tilde{\rho}^{\prime0}$}
\Text(11,-9)[]{\small $\tilde{\rho}^0$}
\Text(-17,-8)[]{\small$\lambda^3_A$}
\Text(-20,-20)[]{\small$\lambda^8_A$} \Text(0,-28)[]{\small
$\rho^{0*}$} \Text(0,30)[]{$\tilde{l}_{L_\alpha}$}
\Text(-50,40)[]{$(b)$}
\end{picture}
\end{center}
\begin{center}
\begin{picture}(120,80)(-60,-40)
\ArrowLine(-50,0)(-34,0) \ArrowLine(-17,0)(-34,0)
\ArrowLine(-17,0)(0,0) \ArrowLine(17,0)(0,0)
\ArrowLine(17,0)(34,0) \ArrowLine(50,0)(34,0)
\DashArrowArcn(0,-20)(40,150,30){4} \DashArrowLine(0,0)(0,-20){3}
\Text(-50,-10)[]{$\mu$} \Text(50,-8)[]{$\mu^c$}
\Text(-50,-18)[]{$(\tau)$} \Text(50,-18)[]{$(\tau^c)$}
\Text(29,-9)[]{\small $\tilde{\rho}_1^{\prime-}$}
\Text(11,-9)[]{\small $\tilde{\rho}_1^{+}$} \Text(-9,-7)[]{\small
$\tilde{W}^-$} \Text(-27,-7)[]{\small $\tilde{W}^+$}
\Text(0,-28)[]{\small $\rho^{0*}$}
\Text(0,30)[]{$\tilde{\nu}_{L\alpha}$} \Text(-50,40)[]{$(c)$}
\end{picture}
\hglue 2.5cm
\begin{picture}(120,80)(-60,-40)
\ArrowLine(-50,0)(-34,0) \ArrowLine(-17,0)(-34,0)
\ArrowLine(-17,0)(0,0) \ArrowLine(17,0)(0,0)
\ArrowLine(17,0)(34,0) \ArrowLine(50,0)(34,0)
\DashArrowArcn(0,-20)(40,150,30){4} \DashArrowLine(0,0)(0,-20){3}
\Text(-50,-10)[]{$\mu $} \Text(50,-8)[]{$\mu^c $}
\Text(-50,-18)[]{$(\tau)$} \Text(50,-18)[]{$(\tau^c)$}
\Text(29,-9)[]{\small $\tilde{\rho}_2^{\prime-}$}
\Text(11,-9)[]{\small $\tilde{\rho}_2^{+}$} \Text(-9,-7)[]{\small
$\tilde{Y}^-$} \Text(-27,-7)[]{\small $\tilde{Y}^+$}
\Text(0,-28)[]{\small $\rho^{0*}$}
\Text(0,30)[]{$\tilde{\nu}_{R\alpha}$} \Text(-50,40)[]{$(d)$}
\end{picture}
\end{center}
\begin{center}
\begin{picture}(120,80)(-60,-40)
\ArrowLine(-50,0)(-34,0) \ArrowLine(0,0)(-34,0)
\ArrowLine(0,0)(34,0) \ArrowLine(50,0)(34,0)
\DashArrowArcn(0,-20)(40,150,90){4}
\DashArrowArcn(0,-20)(40,90,30){4} \DashArrowLine(0,20)(0,40){3}
\Text(-50,-10)[]{$\mu$} \Text(50,-8)[]{$\mu^c$}
\Text(-50,-18)[]{$(\tau)$} \Text(50,-18)[]{$(\tau^c)$}
\Text(15,-8)[]{\small $\tilde{\rho}^{\prime-}_2$}
\Text(-15,-8)[]{\small $\tilde{\rho}^{+}_2$} \Text(10,35)[]{\small
$\rho^{0*}$} \Text(-34,20)[]{$\tilde{\nu}_{L_\alpha}$}
\Text(34,20)[]{$\tilde{\nu}_{R_\beta}$} \Text(-50,40)[]{$(e)$}
\end{picture}
\hglue 2.5cm
\begin{picture}(120,80)(-60,-40)
\ArrowLine(-50,0)(-34,0) \ArrowLine(0,0)(-34,0)
\ArrowLine(0,0)(34,0) \ArrowLine(50,0)(34,0)
\DashArrowArcn(0,-20)(40,150,90){4}
\DashArrowArcn(0,-20)(40,90,30){4} \DashArrowLine(0,20)(0,40){3}
\Text(-50,-10)[]{$\mu$} \Text(50,-8)[]{$\mu^c$}
\Text(-50,-18)[]{$(\tau)$} \Text(50,-18)[]{$(\tau^c)$}
\Text(15,-8)[]{\small $\tilde{\rho}^{\prime-}_1$}
\Text(-15,-8)[]{\small $\tilde{\rho}^{+}_1$} \Text(10,35)[]{\small
$\rho^{0*}$} \Text(-34,20)[]{$\tilde{\nu}_{R_\alpha}$}
\Text(34,20)[]{$\tilde{\nu}_{L_\beta}$} \Text(-50,40)[]{$(f)$}
\end{picture}
\end{center}
\begin{center}
%
\begin{picture}(120,80)(-60,-40)
\ArrowLine(-50,0)(-34,0) \ArrowLine(-17,0)(-34,0)
\ArrowLine(-17,0)(0,0) \ArrowLine(17,0)(0,0)
\ArrowLine(17,0)(34,0) \ArrowLine(50,0)(34,0)
\DashArrowArcn(0,-20)(40,150,30){4} \DashArrowLine(0,0)(0,-20){3}
\Text(-50,-10)[]{$\mu $} \Text(50,-8)[]{$\mu^c $}
\Text(-50,-18)[]{$(\tau)$} \Text(50,-18)[]{$(\tau^c)$}
\Text(-27,-9)[]{\small $\tilde{\rho}^{\prime0}$}
\Text(-11,-9)[]{\small $\tilde{\rho}^0$} \Text(17,-8)[]{\small
$\lambda_B$} \Text(0,-28)[]{\small $\rho^{0*}$}
\Text(0,30)[]{$\tilde{l}_{R_\alpha}$} \Text(-50,40)[]{$(i)$}
\end{picture}
\hglue 2.5cm
\begin{picture}(120,80)(-60,-40)
\ArrowLine(-50,0)(-34,0) \ArrowLine(0,0)(-34,0)
\ArrowLine(0,0)(34,0) \ArrowLine(50,0)(34,0)
\DashArrowArcn(0,-20)(40,150,90){4}
\DashArrowArcn(0,-20)(40,90,30){4} \DashArrowLine(0,20)(0,40){3}
\Text(16,25)[]{$Y_{\mu(\tau)}$}
 \Text(-50,-10)[]{$\mu$}
\Text(50,-8)[]{$\mu^c$} \Text(0,-8)[]{\small $\lambda_B$}
\Text(10,40)[]{\small $\rho^{*0}$}
\Text(-34,20)[]{$\tilde{l}_{L_\alpha}$}
\Text(36,18)[]{$\tilde{l}_{R_\beta}$} \Text(-50,40)[]{$(k)$}
\Text(-50,-18)[]{$(\tau)$} \Text(50,-18)[]{$(\tau^c)$}
\end{picture}
\end{center}

\vspace{-0.3cm} \caption{\small Diagrams  contributing to
$\Delta^{1\rho}_{\mu}$ ( or $  \Delta^{\rho}_{\tau})$.}
\label{Drhomu1}
\end{figure}
\begin{figure}[t]
\vspace{0.4cm}
\begin{center}
\begin{picture}(120,80)(-60,-40)
\ArrowLine(-50,0)(-34,0) \ArrowLine(0,0)(-34,0)
\ArrowLine(0,0)(34,0) \ArrowLine(50,0)(34,0)
\DashArrowArcn(0,-20)(40,150,90){4}
\DashArrowArcn(0,-20)(40,90,30){4} \DashArrowLine(0,20)(0,40){3}
\Text(10,25)[]{$Y_{\tau}$} \Text(-50,-10)[]{$\mu$}
\Text(50,-8)[]{$\mu^c$} \Text(0,-8)[]{\small $\lambda_B$}
\Text(10,40)[]{\small $\rho^{0*}$}
\Text(-34,20)[]{$\tilde{l}_{L_\alpha}$}
\Text(34,20)[]{$\tilde{l}_{R_\beta}$} 
\end{picture}
\end{center}
\vspace{-0.3cm} \caption{\small Diagram contributing to
 $\Delta^{2\rho}_\mu$.} \label{Delrhop}
\end{figure}
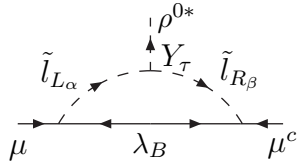
\begin{figure}[t]
\vspace{0.4cm}
\begin{center}
\begin{picture}(120,80)(-60,-40)
\ArrowLine(-50,0)(-34,0) \ArrowLine(0,0)(-34,0)
\ArrowLine(0,0)(34,0) \ArrowLine(50,0)(34,0)
\DashArrowArcn(0,-20)(40,150,90){4}
\DashArrowArcn(0,-20)(40,90,30){4} \DashArrowLine(0,40)(0,20){3}
\Text(-50,-10)[]{$\mu$} \Text(50,-8)[]{$\mu^c$}
\Text(-50,-18)[]{$(\tau)$} \Text(50,-18)[]{$(\tau^c)$}
\Text(15,-8)[]{\small $\tilde{\rho}^{\prime-}_2$}
\Text(-15,-8)[]{\small $\tilde{\rho}^{+}_2$} \Text(10,35)[]{\small
$\rho^{\prime0}$} \Text(-34,20)[]{$\tilde{\nu}_{L_\alpha}$}
\Text(34,20)[]{$\tilde{\nu}_{R_\beta}$} \Text(-50,40)[]{$(a)$}
\end{picture}
\hglue 2.5cm
\begin{picture}(120,80)(-60,-40)
\ArrowLine(-50,0)(-34,0) \ArrowLine(0,0)(-34,0)
\ArrowLine(0,0)(34,0) \ArrowLine(50,0)(34,0)
\DashArrowArcn(0,-20)(40,150,90){4}
\DashArrowArcn(0,-20)(40,90,30){4} \DashArrowLine(0,40)(0,20){3}
\Text(-50,-10)[]{$\mu$} \Text(50,-8)[]{$\mu^c$}
\Text(-50,-18)[]{$(\tau)$} \Text(50,-18)[]{$(\tau^c)$}
\Text(15,-8)[]{\small $\tilde{\rho}^{\prime-}_1$}
\Text(-15,-8)[]{\small $\tilde{\rho}^{+}_1$} \Text(10,35)[]{\small
$\rho^{\prime0}$} \Text(-34,20)[]{$\tilde{\nu}_{R_\alpha}$}
\Text(34,20)[]{$\tilde{\nu}_{L_\beta}$} \Text(-50,40)[]{$(b)$}
\end{picture}
\end{center}
\begin{center}
\begin{picture}(120,80)(-60,-40)
\ArrowLine(-50,0)(-34,0) \ArrowLine(0,0)(-34,0)
\ArrowLine(0,0)(34,0) \ArrowLine(50,0)(34,0)
\DashArrowArcn(0,-20)(40,150,90){4}
\DashArrowArcn(0,-20)(40,90,30){4} \DashArrowLine(0,40)(0,20){3}
\Text(-50,-10)[]{$\mu$} \Text(50,-8)[]{$\mu^c$}
\Text(-50,-18)[]{$(\tau)$} \Text(50,-18)[]{$(\tau^c)$}
\Text(0,-8)[]{\small $\lambda_B$} \Text(10,35)[]{\small
$\rho^{\prime0}$} \Text(-34,20)[]{$\tilde{l}_{L_\alpha}$}
\Text(34,20)[]{$\tilde{l}_{R_\beta}$} \Text(-50,40)[]{$(c)$}
\end{picture}
\end{center}
\vspace{-0.3cm} \caption{\small Diagrams  contributing to
 $\Delta^{1\rho'}_\mu$ $[(a), (b)]$, $\Delta^{2\rho'}_\mu$ $[ (c)]$
 ( or $\Delta^{\rho'}_\tau$  $[(a), (b), (c)]$).} \label{Delrhop1}
\end{figure}
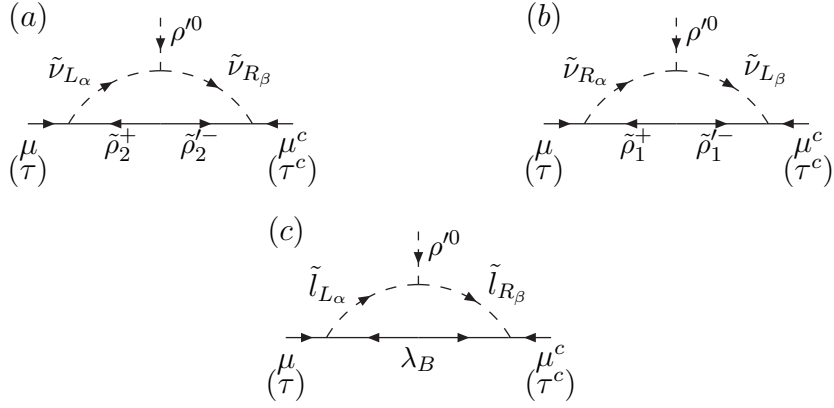
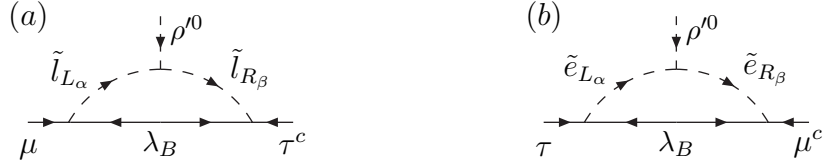
\begin{figure}[t]
\vspace{0.4cm}
\begin{center}
\begin{picture}(120,80)(-60,-40)
\ArrowLine(-50,0)(-34,0) \ArrowLine(0,0)(-34,0)
\ArrowLine(0,0)(34,0) \ArrowLine(50,0)(34,0)
\DashArrowArcn(0,-20)(40,150,90){4}
\DashArrowArcn(0,-20)(40,90,30){4} \DashArrowLine(0,40)(0,20){3}
\Text(-50,-10)[]{$\mu$} \Text(50,-8)[]{$\tau^c$}
\Text(0,-8)[]{\small $\lambda_B$} \Text(10,35)[]{\small
$\rho^{\prime0}$} \Text(-34,20)[]{$\tilde{l}_{L_\alpha}$}
\Text(34,20)[]{$\tilde{l}_{R_\beta}$} \Text(-50,40)[]{$(a)$}
\end{picture}
\hglue 2.5cm
\begin{picture}(120,80)(-60,-40)
\ArrowLine(-50,0)(-34,0) \ArrowLine(0,0)(-34,0)
\ArrowLine(0,0)(34,0) \ArrowLine(50,0)(34,0)
\DashArrowArcn(0,-20)(40,150,90){4}
\DashArrowArcn(0,-20)(40,90,30){4} \DashArrowLine(0,40)(0,20){3}
\Text(-50,-10)[]{$\tau$} \Text(50,-8)[]{$\mu^c$}
\Text(0,-8)[]{\small $\lambda_B$} \Text(10,35)[]{\small
$\rho^{\prime0}$} \Text(-34,20)[]{$\tilde{e}_{L_\alpha}$}
\Text(34,20)[]{$\tilde{e}_{R_\beta}$} \Text(-50,40)[]{$(b)$}
\end{picture}
\end{center}
\vspace{-0.3cm} \caption{\small Diagrams  contributing to
$\Delta^{\rho'}_L$ [(a)] and $\Delta^{\rho'}_R $ [(b)].}
\label{rhop}
\end{figure}

\section{\label{Lagrangian1}Lagrangian}
We  have
 denoted  by  $f_{L,R} (\bar{f}_{L,R}^c)$ two
component spinor of the generic  matter left-handed and
right-handed fermion, respectively. The $\tilde{f}_L
(\tilde{f}_L^c)$ are their superpartners which satisfy
$(\bar{f}_L^c= (f_R)^{\dagger T}, \tilde{f}_L^c=\tilde{f}^*_R $).
The four-component Dirac spinor can be represented through
two-component spinor such as:
$\mu\equiv(\mu_L,\;\mu_R)=(\mu_L,\;\bar{\mu}_L^c)$ and
$\widetilde{{\mu}}\equiv(\tilde{\mu}_L,\;\tilde{\mu}_R)=
(\widetilde{\mu}_L,\;\widetilde{\mu}_L^{c*})$. We also emphasize
that three-left handed leptons contained in the triplet $L_{aL}$
of $\mathrm{SU}(3)_L$ are $(f_{a1L}, f_{a2L},f_{a3L})$: \bea
f_{a1L}&\in&\{\nu_{eL},\nu_{\mu L},\nu_{\tau
L}\}\equiv\{\nu_{1L},\nu_{2L},\nu_{3L}\},\crn
f_{a2L}&\in&\{e_L,\mu_L,\tau_L\},\crn f_{a3L} &\in&
\{\nu^c_{eL},\nu^c_{\mu L},\nu^c_{\tau
L}\}\equiv\{\nu^c_{1L},\nu^c_{2L},
\nu^c_{3L}\}\equiv\{(\nu^c_1)_L,(\nu^c_2)_L, (\nu^c_3)_L\}
\label{f123} \eea
 while $f_{aL}^c$ is singlet
under  $\mathrm{SU}(3)_L$. Conventions for two component spinors
used in our paper are the same as those given in \cite{Anna2}
except the lower index $L$, which is used to distinguish between
Dirac spinors $f$ and left-handed Weyl spinors $f_L$.

 Next, let us find the interactional vertices  relating with
  our calculation. We start to collect related terms from the
 Lagrangian given in  ~\cite{Dong1,s331r} into the following ones:
\begin{enumerate}
    \item  Gaugino and Higgsino mass terms:
\bea \mathcal{L}_{gh} &=&-  \left[ \frac{1}{2} m_{ \lambda}
\sum_{b=1}^{8} \left( \lambda^{b}_{A} \lambda^{b}_{A} \right)  +
\frac{1}{2}  m^{ \prime} \lambda_{B} \lambda_{B} + \mu_{\chi}
\widetilde{\chi} \widetilde{\chi}^\prime
               +\mu_{\rho}\widetilde{ \rho}
               \widetilde{\rho}^\prime\right]+ \mathrm{H.c.}\crn
               &=& -\frac{1}{2}m^{ \prime} \lambda_{B} \lambda_{B}
               -m_{\lambda}\tilde{W}^+\tilde{W}^- -\frac{1}{2}
               m_{\lambda}\lambda_A^3\lambda_A^3\crn&-&m_{\lambda}\tilde{Y}^{\prime+}
               \tilde{Y}^{\prime-}-m_{\lambda}\tilde{X^{0*}}
               \tilde{X^0}-\frac{1}{2}m_{\lambda}\lambda_A^8\lambda_A^8
-\mu_{\chi}\left( \widetilde{\chi}^0_1 \widetilde{\chi}^{\prime
0}_1+\widetilde{\chi}^- \widetilde{\chi}^{\prime
+}+\widetilde{\chi}^0_2 \widetilde{\chi}^{\prime 0}_2\right) \crn
&-& \mu_{\rho}\left( \widetilde{\rho}^+_1 \widetilde{\rho}^{\prime
-}_1+\widetilde{\rho}^0 \widetilde{\rho}^{\prime
0}+\widetilde{\rho}^+_2 \widetilde{\rho}^{\prime -}_2\right)+
\mathrm{H.c.}.\label{GHmass}\eea where \bea \tilde{W}^\pm
&\equiv&\frac{1}{\sqrt{2}}(\lambda_A^1\mp i \lambda_A^2), \hs
\tilde{Y}^{\pm}\equiv\frac{1}{\sqrt{2}}(\lambda_A^6\pm i
\lambda_A^7) ,\crn
 \tilde{X}&\equiv& \frac{1}{\sqrt{2}}(\lambda_A^4 +i \lambda_A^5),
 \hs \tilde{X}^{*}\equiv \frac{1}{\sqrt{2}}(\lambda_A^4 -i \lambda_A^5) \eea
where we have used the results of mass eigenstate states for Higgsinos
and gauginos  given  in \cite{Dong2} and \cite{Long2}.
 \item Fermion-sfermion-gaugino interaction terms:
   \begin{eqnarray}
\mathcal{L}_{l \tilde{l}
\tilde{V}}&=&-\frac{ig^{\prime}}{\sqrt{3}}\left[
-\frac{1}{3}(\bar{L}\tilde{L}\bar{\lambda}_{B}-
\tilde{L}^{\dagger}L\lambda_{B})+
(\bar{l}^c\tilde{l}^c\bar{\lambda}_{B}-
\tilde{l}^{c*}l^c\lambda_{B}) \right]\crn&-& \frac{ig}{\sqrt{2}} (
\bar{L}\lambda^i\tilde{L}\bar{\lambda}^i_{A}-
\tilde{L}^{\dagger}\lambda^iL\lambda^i_{A}),\label{Ll-sl-gaugino1}
\end{eqnarray}
where  $i=1,2,...,8$ is a color index and
$L\equiv(L_{1L},\;L_{2L},\;L_{3L})^T$, $\tilde{L}\equiv
(\tilde{L}_{1L},\; \tilde{L}_{2L},\;\tilde{L}_{3L})^T$,
$\bar{l}^c=(l_{1R},\;l_{2R},\;
l_{3R})^T\equiv(\bar{l}^c_{1L},\;\bar{l}^c_{2L},\;
\bar{l}^c_{3L})^T\equiv(e_R,\mu_R,\tau_R)^T\equiv
(\bar{e}^c_L,\;\bar{\mu}^c_L,\;\bar{\tau}^c_L)^T$, and
$\tilde{l}^c=(\tilde{l}^*_{1R},\;\tilde{l}^*_{2R},\;
\tilde{l}^*_{3R})^T\equiv (\tilde{e}^*_R,\;
\tilde{\mu}^*_R,\;\tilde{\tau}^*_R,)^T$.
In this  paper we just focus on interactions relating with two
fermions, namely $l_{aL}=\{\mu_L,\tau_L\}$.  All interested terms
are given as
\bea \mathcal{L}_{l \tilde{l}
\tilde{V}}&=&\left[\bar{\mu}_{L}\left(\frac{ig'}
{3\sqrt{3}}\bar{\lambda}_B+
\frac{ig}{\sqrt{2}}(\bar{\lambda}^3_A-\frac{1}{\sqrt{3}}
\bar{\lambda}^8_A)\right)\tilde{\mu }_{L}\right.\crn
&-&\left.\mu_{L}\left(\frac{ig'}{3\sqrt{3}}\lambda_B+
\frac{ig}{\sqrt{2}}(\lambda^3_A
-\frac{1}{\sqrt{3}}\lambda^8_A)\right)\tilde{\mu}^*_{L} \right.
\crn
&-& \left. \frac{ig'}{\sqrt{3}}(\bar{\mu}^c_{L}\tilde{\mu}^c_{L}
\bar{\lambda}_B-\mu^c_{L}\tilde{\mu}^{c*}_{L}\lambda_B) +
\frac{ig}{\sqrt{2}}\left(
\bar{\mu}_L\tilde{\mu}_L\bar{\lambda}^3_A- \tilde{\mu}^*_L
\mu_L\lambda^3_A\right) \right.\crn
&+& \left.\frac{ig}{\sqrt{2}}\left(
\bar{\mu}_L\tilde{\mu}_L\bar{\lambda}^8_A- \tilde{\mu}^*_L
\mu_L\lambda^8_A\right)\right.\crn
&-& \left. ig\left((\bar{\mu
}_{L}\overline{\tilde{W}}^{+}\tilde{\nu}_{\mu
L}-\mu_{L}\tilde{W}^{+}\tilde{\nu}^*_{\mu L})+
(\bar{\mu}_{L}\overline{\tilde{Y}}^{+}\tilde{\nu}^c_{\mu
L}-\mu_{L}\tilde{Y}^{+}\tilde{\nu}^{c*}_{\mu L})\right)\right]\crn
&+& [\mu\rightarrow\tau]. \label{Ll-sl-gaugino3}\eea
 From this, we list the
related vertices in Table \ref{tree-vertex1}
\begin{table}
  \centering
  \begin{tabular}{|c|c|c|c|}
    \hline
    Vertex & Factor &Vertex & Factor \\
    \hline
      $\bar{\mu}_L\bar{\lambda}_B\tilde{\mu}_L$& $-\frac{g'}{3\sqrt{3}}$
      &$\mu_L\lambda_B\tilde{\mu}^*_L$ & $\frac{g'}{3\sqrt{3}}$ \\
$\bar{\mu}^c_{L}\tilde{\mu}^c_{L} \bar{\lambda}_B $ &
$\frac{g'}{\sqrt{3}}$ & $\mu^c_{L}\tilde{\mu}^{c*}_{L}\lambda_B $
& $\frac{-g'}{\sqrt{3}} $  \\
  $\bar{\mu}_L\tilde{\mu}_L\bar{\lambda}^3_A$ & $\frac{-g}{\sqrt{2}}$
  & $\tilde{\mu}^*_L
\mu_L\lambda^3_A$ & $\frac{g}{\sqrt{2}}$ \\
 $\bar{\mu}_L\tilde{\mu}_L\bar{\lambda}^8_A$ & $
  \frac{g}{\sqrt{6}}$ & $\tilde{\mu}^*_L
\mu_L\lambda^8_A $ & $ \frac{-g}{\sqrt{6}}$  \\
 $\bar{\mu
}_{L}\overline{\tilde{W}}^{+}\tilde{\nu}_{\mu L}$  & $g$ &
$\mu_{L}\tilde{W}^{+}\tilde{\nu}^*_{\mu L}$ & $-g$ \\
$\bar{\mu}_{L}\overline{\tilde{Y}}^{+}\tilde{\nu}^c_{\mu
L}$ & $g$ & $\mu_{L}\tilde{Y}^{+}\tilde{\nu}^{c*}_{\mu L}$ & $-g$ \\
$\bar{\tau}_L\bar{\lambda}_B\tilde{\tau}_L$&
$-\frac{g'}{3\sqrt{3}}$
      &$\tau_L\lambda_B\tilde{\tau}^*_L$ & $\frac{g'}{3\sqrt{3}}$ \\
$\bar{\tau}^c_{L}\tilde{\tau}^c_{L} \bar{\lambda}_B $ &
$\frac{g'}{\sqrt{3}}$ & $\tau^c_{L}\tilde{\tau}^{c*}_{L}\lambda_B
$
& $\frac{-g'}{\sqrt{3}} $  \\
  $\bar{\tau}_L\tilde{\tau}_L\bar{\lambda}^3_A$ & $\frac{-g}{\sqrt{2}}$
  & $\tilde{\tau}^*_L
\tau_L\lambda^3_A$ & $\frac{g}{\sqrt{2}}$ \\
 $\bar{\tau}_L\tilde{\tau}_L\bar{\lambda}^8_A$ & $ \frac{g}
 {\sqrt{6}}$ & $\tilde{\tau}^*_L
\tau_L\lambda^8_A $ & $ \frac{g}{-\sqrt{6}}$  \\
 $\bar{\tau
}_{L}\overline{\tilde{W}}^{+}\tilde{\nu}_{\tau L}$  & $g$ &
$\tau_{L}\tilde{W}^{+}\tilde{\nu}^*_{\tau L}$ & $-g$ \\
$\bar{\tau}_{L}\overline{\tilde{Y}}^{+}\tilde{\nu}^c_{\tau
L}$ & $g$ & $\tau_{L}\tilde{Y}^{+}\tilde{\nu}^{c*}_{\tau L}$ & $-g$ \\
    \hline
  \end{tabular}
  \caption{Vertices of  lepton-slepton-gaugino interaction at
  tree level.}\label{tree-vertex1}
\end{table}
 \item Higgs-Higgsino-gaugino:
 \bea
   \mathcal{L}_{H \tilde{H} \tilde{V}}&=&
   -\frac{ig^{ \prime}}{ \sqrt{3}} \left[
- \frac{1}{3} (\bar{\tilde{\chi}} \chi \bar{\lambda}_{B}-
\chi^{\dagger}\tilde{\chi}\lambda_{B}) + \frac{1}{3}
(\bar{\tilde{\chi}}^{\prime} \chi^{\prime} \bar{\lambda}_{B}-
\chi^{\prime\dagger}\tilde{\chi}^{\prime}\lambda_{B}) \right. \crn
&+& \left.\frac{2}{3} (\bar{\tilde{\rho}} \rho \bar{\lambda}_{B}-
\rho^{\dagger}\tilde{\rho}\lambda_{B}) - \frac{2}{3}
(\bar{\tilde{\rho}}^{\prime} \rho^{\prime} \bar{\lambda}_{B}-
\rho^{\prime\dagger}\tilde{\rho}^{\prime}\lambda_{B}) \right]\crn
&-& \frac{ig}{ \sqrt{2}} \left[
\bar{\tilde{\rho}}\lambda^a\rho\bar{\lambda}^a_{A} -
\rho^{\dagger}\lambda^a\tilde{\rho}\lambda^a_{A}+
\bar{\tilde{\chi}}\lambda^a\chi\bar{\lambda}^a_{A} -
\chi^{\dagger}\lambda^a\tilde{\chi}\lambda^a_{A} \right. \crn &-&
\left. \bar{\tilde{\rho}}^{\prime}\lambda^{*
a}\rho^{\prime}\bar{\lambda}^a_{A} +
\rho^{\prime\dagger}\lambda^{*
a}\tilde{\rho}^{\prime}\lambda^a_{A}-
\bar{\tilde{\chi}}^{\prime}\lambda^{*
a}\chi^{\prime}\bar{\lambda}^a_{A} +\chi^{\prime\dagger}\lambda^{*
a}\tilde{\chi}^{\prime}\lambda^a_{A}\right]
\label{hhiggsinogaugino1}\eea Vertices of neutral
Higgs-Higgsino-gaugino interactions are shown in table
\ref{hhiggsinogaugino2}
\begin{table}
  \centering
  \caption{Vertices of the neutral Higgs-Higgsino-gaugino interactions.}\label{hhiggsinogaugino2}
 \begin{tabular}{|c|c|c|c|}
    \hline
    Vertex & Factor & Vertex & Factor \\
    \hline
     $\bar{\tilde{\chi}}^0_1\chi^0_1\bar{\lambda}_B $ & $\frac{-g'}{3\sqrt{3}}$
     & $\tilde{\chi}^0_1\chi^{0\dagger}_1\lambda_B $  & $\frac{g'}{3\sqrt{3}}$ \\
 $\bar{\tilde{\chi}}^0_2\chi^0_2\bar{\lambda}_B $ & $\frac{-g'}{3\sqrt{3}}$
     & $\tilde{\chi}^0_2\chi^{0\dagger}_2\lambda_B $  & $\frac{g'}{3\sqrt{3}}$ \\
     $\bar{\tilde{\chi}}^0_1\chi^0_1\bar{\lambda}^3_A$ & $\frac{g}{\sqrt{2}}$&
      $\tilde{\chi}^0_1\chi^{0\dagger}_1\lambda^3_A $ &$\frac{-g }{\sqrt{2}}$ \\
     $\bar{\tilde{\chi}}^0_1\chi^0_1\bar{\lambda}^8_A$ & $\frac{g}{\sqrt{6}}$&
      $\tilde{\chi}^0_1\chi^{0\dagger}_1\lambda^8_A $ &$\frac{-g }{\sqrt{6}}$ \\
      $\bar{\tilde{\chi}}^0_2\chi^0_2\bar{\lambda}^8_A$ & $
      \frac{-g \sqrt{2}}{\sqrt{3}}$
       & $\tilde{\chi}^0_2\chi^{0\dagger}_2\lambda^8_A $ &
       $\frac{g \sqrt{2}}{\sqrt{3}}$\\
    $\bar{\tilde{\chi}}^{\prime0}_1\chi^{\prime0}_1\bar{\lambda}_B $
    & $\frac{g'}{3\sqrt{3}}$
     & $\tilde{\chi}^{\prime0}_1\chi^{\prime0\dagger}_1\lambda_B $
     & $\frac{-g'}{3\sqrt{3}}$ \\
 $\bar{\tilde{\chi}}^{\prime0}_2\chi^{\prime0}_2\bar{\lambda}_B $
 & $\frac{g'}{3\sqrt{3}}$
     & $\tilde{\chi}^{\prime0}_2\chi^{\prime0\dagger}_2\lambda_B $
      & $\frac{-g'}{3\sqrt{3}}$ \\
     $\bar{\tilde{\chi}}^{\prime0}_1\chi^{\prime0}_1
     \bar{\lambda}^3_A$ & $\frac{-g }{\sqrt{2}}$&
      $\tilde{\chi}^{\prime0}_1\chi^{\prime0\dagger}_1\lambda^3_A $
      &$\frac{g }{\sqrt{2}}$ \\
    $\bar{\tilde{\chi}}^{\prime0}_1\chi^{\prime0}_1
     \bar{\lambda}^8_A$ & $\frac{-g }{\sqrt{6}}$&
      $\tilde{\chi}^{\prime0}_1\chi^{\prime0\dagger}_1\lambda^8_A $
      &$\frac{g }{\sqrt{6}}$ \\
      $\bar{\tilde{\chi}}^{\prime0}_2\chi^{\prime0}_2\bar{\lambda}^8_A$
      & $\frac{g \sqrt{2}}{\sqrt{3}}$
       & $\tilde{\chi}^{\prime0}_2\chi^{\prime0\dagger}_2\lambda^8_A $
       & $\frac{-g \sqrt{2}}{\sqrt{3}}$\\
    $\bar{\tilde{\rho}}^0\rho^0\bar{\lambda}_B $ & $\frac{2g'}{3\sqrt{3}}$
     & $\tilde{\rho}^0\rho^{0\dagger}\lambda_B $  & $\frac{-2g'}{3\sqrt{3}}$ \\
     $\bar{\tilde{\rho^0}}\rho^0\bar{\lambda}^3_A$ & $\frac{-g }{\sqrt{2}}$&
      $\tilde{\rho}^0\rho^{0\dagger}\lambda^3_A $ &$\frac{g}{\sqrt{2}}$ \\
     $\bar{\tilde{\rho}}^0\rho^0\bar{\lambda}^8_A$ & $\frac{g }{\sqrt{6}}$&
      $\tilde{\rho}^0\rho^{0\dagger}\lambda^8_A $ &$\frac{-g }{\sqrt{6}}$ \\
$\bar{\tilde{\rho}}^{\prime0}\rho^{\prime0}\bar{\lambda}_B $ &
$\frac{-2g'}{3\sqrt{3}}$
     & $\tilde{\rho}^{\prime0}\rho^{\prime0\dagger}\lambda_B $
     & $\frac{2g'}{3\sqrt{3}}$ \\
     $\bar{\tilde{\rho^{\prime0}}}\rho^{\prime0}\bar{\lambda}^3_A$
     & $\frac{g }{\sqrt{2}}$&
      $\tilde{\rho}^{\prime0}\rho^{\prime0\dagger}\lambda^3_A $
      &$\frac{-g}{\sqrt{2}}$ \\
     $\bar{\tilde{\rho}}^{\prime0}\rho^{\prime0}\bar{\lambda}^8_A$
      & $\frac{-g }{\sqrt{6}}$&
      $\tilde{\rho}^{\prime0}\rho^{\prime0\dagger}\lambda^8_A $
      &$\frac{g }{\sqrt{6}}$ \\
 $\bar{\tilde{\chi}}^-\bar{\tilde{W}}^+\chi^0_1$& $g$&
 $\chi^{0*}\tilde{W}^+\tilde{\chi}^-$& -$g$\\
$\bar{\tilde{\chi}}^-\bar{\tilde{Y}}^+\chi^0_2$& $g$&
$\chi^{0*}_2\tilde{Y}^+\tilde{\chi}^-$& $-g$\\
   $\bar{\tilde{\chi}}^0_2\bar{\tilde{X}}\chi^0_1$&
   $g$&$\chi^{0*}_1\tilde{X}\tilde{\chi}^0_2$& $-g$\\
 $\bar{\tilde{\chi}}^0_1\bar{\tilde{X}}^*\chi^0_2$&
 $g$&$\chi^{0*}_2\tilde{X}^*\tilde{\chi}^0_1$& $-g$\\
$\bar{\tilde{\rho}}^+_1\bar{\tilde{W}}^-\rho^0$& $g$&
 $\rho^{0*}\tilde{W}^-\tilde{\rho}^+_1$& -$g$\\
$\bar{\tilde{\rho}}^+_2\bar{\tilde{Y}}^-\rho^0$& $g$&
 $\rho^{0*}\tilde{Y}^-\tilde{\rho}^+_2$& -$g$\\
 $\bar{\tilde{\chi}}^{\prime+}\bar{\tilde{W}}^-\chi^{\prime0}_1$& $-g$&
 $\chi^{0*}_1\tilde{W}^-\tilde{\chi}^{\prime+}$& $g$\\
$\bar{\tilde{\chi}}^{\prime+}\bar{\tilde{Y}}^-\chi^{\prime0}_2$&
$-g$&
$\chi^{\prime0*}_2\tilde{Y}^-\tilde{\chi}^{\prime+}$& $g$\\
   $\bar{\tilde{\chi}}^{\prime0}_2\bar{\tilde{X}}^*\chi^{\prime0}_1$&
   $-g$&$\chi^{\prime0*}_1\tilde{X}\tilde{\chi}^{\prime0}_2$& $g$\\
 $\bar{\tilde{\chi}}^{\prime0}_1\bar{\tilde{X}}\chi^{\prime0}_2$&
 $-g$&$\chi^{\prime0*}_2\tilde{X}\tilde{\chi}^{\prime0}_1$& $g$\\
$\bar{\tilde{\rho}}^{\prime-}_1\bar{\tilde{W}}^+\rho^{\prime0}$&
$-g$&
 $\rho^{\prime0*}\tilde{W}^+\tilde{\rho}^{\prime-}_1$& $g$\\
$\bar{\tilde{\rho}}^{\prime-}_2\bar{\tilde{Y}}^+\rho^{\prime0}$&
$-g$&
 $\rho^{\prime0*}\tilde{Y}^+\tilde{\rho}^{\prime-}_2$& $g$\\

    \hline
  \end{tabular}
\end{table}

\item \textbf{Yukawa interaction terms}:
    \begin{eqnarray}
\mathcal{L}_{l \tilde{l} \tilde{H}}&=&- \frac{ \lambda_{1ab}}{3}
\left( L_{aL} \tilde{\rho}^{\prime} \tilde{l}^{c}_{bL}+
\tilde{L}_{aL} \tilde{\rho}^{\prime}l^{c}_{bL} \right)- \frac{
\lambda_{3ab}}{3} \left( L_{aL} \tilde{ \rho}
\tilde{L}_{bL}+\tilde{L}_{aL} \tilde{ \rho}L_{bL} \right), \crn
\label{lslH}\\
\mathcal{L}_{llH}&=&- \frac{ \lambda_{1ab}}{3} L_{aL}l^{c}_{bL}
\rho^{\prime}- \frac{
\lambda_{3ab}}{3}\epsilon^{\alpha\beta\gamma} (L_{aL})_{\alpha}
(L_{bL})_{\beta} (\rho)_{\gamma}+ \mathrm{H.c.}. \label{lepbos}
\end{eqnarray}

Our work needs only  terms which include leptons $\mu$ or $\tau$,
such as:
\bea \mathcal{L}_{l\tilde{l}\tilde{H}}&=& - \frac{
\lambda_{1ab}}{3} \left[l_{aL}\tilde{\rho}^{\prime
0}\tilde{l}^c_{bL}+(\tilde{\nu}_{aL}\tilde{\rho}^{\prime
-}_1+\tilde{l}_{aL}\tilde{\rho}^{\prime
0}+\tilde{\nu}^c_{aL}\tilde{\rho}^{\prime -}_2)l^c_{bL}\right]\crn
&-&\frac{ \lambda_{3ab}}{3}
\left[l_{aL}\tilde{\rho}^{+}_2\widetilde{\nu}_{bL}
-l_{aL}\widetilde{\rho}^{+}_1\widetilde{\nu}^c_{bL}+
\widetilde{\nu}^c_{aL}\widetilde{\rho}^{+}_1l_{bL}-
\widetilde{\nu}_{aL}\widetilde{\rho}^{+}_2l_{bL} \right],
\label{lslhiggino1}\eea

From now we just note that because the conversation of lepton
flavor in the lepton sector at tree level then $\lambda_{1ab}=0$
with $a \neq b$   and $\lambda_{3cd}=0$ with $c = d$. For
simplicity, we use new notations: $Y_{e}\equiv\lambda_{111}/3,
Y_{\mu}\equiv\lambda_{122}/3, Y_{\tau}\equiv\lambda_{133}/3 $ and
$Y_{\nu_{e\mu}}\equiv\lambda_{312}/3,
Y_{\nu_{\mu\tau}}\equiv\lambda_{323}/3,
Y_{\nu_{e\tau}}\equiv\lambda_{313}/3 $.  Eq.(\ref{lslhiggino1})
now can be written in the common form:
   \bea
   \mathcal{L}_{l\tilde{l}\tilde{H}}&=&
   -Y_{\mu}\left[\mu_L\tilde{\mu}^c_L\tilde{\rho}^{\prime0}+(\tilde{\nu}_{\mu
   L}\tilde{\rho}^{\prime -}_1+\tilde{\mu}_{L}\tilde{\rho}^{\prime
   0}+\tilde{\nu}^c_{\mu L}\tilde{\rho}^{\prime -}_2)\mu^c_{L}\right]\crn
&-&
Y_{\tau}\left[\tau_L\tilde{\tau}^c_L\tilde{\rho}^{\prime0}+(\tilde{\nu}_{\tau
L}\tilde{\rho}^{\prime -}_1+\tilde{\tau}_{L}\tilde{\rho}^{\prime
0}+\tilde{\nu}^c_{\tau L}\tilde{\rho}^{\prime
-}_2)\tau^c_{L}\right]\crn
&-& Y_{\nu_{ab}}[l_{aL}\tilde{\rho}^{+}_2\tilde{\nu}_{bL}
-l_{aL}\tilde{\rho}^{+}_1\tilde{\nu}^c_{bL}+
\tilde{\nu}^c_{aL}\tilde{\rho}^{+}_1l_{bL}-
\tilde{\nu}_{aL}\tilde{\rho}^{+}_2l_{bL}]
   \label{lslhiggino2}\eea

Corresponding vertices are shown in Fig. \ref{lslhiggino3}
\begin{table}
\begin{tabular}{|c|c|c|c|}
  \hline
  Vertex & Factor & vertex & Factor \\
  \hline
  $\mu_L\tilde{\mu}^c_L\tilde{\rho}^{\prime0}$ & $-iY_{\mu}$  & $\tilde{\nu}_{\mu
   L}\tilde{\rho}^{\prime -}_1 \mu^c_{L}$ & $-iY_{\mu}$  \\
 $\tilde{\mu}_{L}\tilde{\rho}^{\prime
   0} \mu^c_{L}$ & $-iY_{\mu}$ &$\tilde{\nu}^c_{\mu L}\tilde{\rho}^{\prime
   -}_2\mu^c_{L}$& $-iY_{\mu}$\\
  $\tau_L\tilde{\tau}^c_L\tilde{\rho}^{\prime0}$ & $-iY_{\tau}$  & $\tilde{\nu}_{\tau
   L}\tilde{\rho}^{\prime -}_1\tau^c_{L}$ & $-iY_{\tau}$  \\
 $\tilde{\tau}_{L}\tilde{\rho}^{\prime
   0} \tau^c_{L}$ & $-iY_{\tau}$ &$\tilde{\nu}^c_{\tau L}\tilde{\rho}^{\prime
   -}_2\tau^c_{L}$& $-iY_{\tau}$\\
   $\mu_{L}~\tilde{\rho}^{+}_2~\tilde{\nu}_{\tau_L}$& $
   -2iY_{\nu_{\mu\tau}}$
   &$\mu_{L}~\tilde{\rho}^{+}_1~\tilde{\nu}^c_{\tau}$&
   $2iY_{\nu_{\mu\tau}}$ \\
$\tau_{L}~\tilde{\rho}^{+}_2~\tilde{\nu}_{\mu_L}$&
$-2iY_{\nu_{\tau\mu}}$
   &$\tau_{L}~\tilde{\rho}^{+}_1~\tilde{\nu}^c_{\mu_L}$
   & $2iY_{\nu_{\tau\mu}}$ \\
  \hline
\end{tabular}
  \centering
  \caption{Higgsino-lepton-slepton interacions}\label{lslhiggino3}
\end{table}
 \item In the soft Lagrangian, the mass term  of sleptons
 is given by

 \bea \mathcal{L}_{\tilde{f}mass}&=&-\tilde{m}^2_{Lab}
 \tilde{L}_{aL}^\dagger \tilde{L}_{bL}
-\tilde{m}^2_{Rab}\tilde{l}_{aL}^{c*}
\widetilde{l}_{bL}^c-\left[h^{\prime}_{ab}\tilde{L}^T_{aL}
\rho^{\prime}\tilde{l}^c_{bL}\right.\crn &+&\left.h_{ab}
\varepsilon^{\alpha\beta\gamma}
(\tilde{L}_{aL})_{\alpha}(\tilde{L}_{bL})_{\beta}(\rho)_{\gamma} +
\frac{\lambda_{1ab}}{3} \mu_{\rho} \rho^* \tilde{L}_{aL}
\tilde{l}^c_{bL} \right.\crn &+&\left. \frac{\lambda_{3ab}}{3}
\mu_{\rho}\varepsilon^{\alpha\beta\gamma} (\rho^{\prime
*})_{\alpha} (\tilde{L}_{aL})_{\beta} (\tilde{L}_{bL})_{\gamma}
+\mathrm{H.c.}\right], \label{sfer-mass}\eea
 here $a, b$  are flavor indices $\{a,b=1,2,3\}$ or
 $a,b=\{e,\mu,\tau\}$ and  $\alpha,\beta, \gamma$ are
 component indices of $SU(3)_L$. The $\varepsilon^{\alpha\beta\gamma}$ is
 the antisymmetric tensor. In this paper we focus on the
 mixing of slepton $\widetilde{\mu}$ and $\widetilde{\tau}$.
 This mixing makes
 mass-eigenstate basis of slepton is different from \cite{Dong2}.
 For more detail, please see in Appendix \ref{eigenstate}. The Lagrangian
 relating with  Higgs-lepton-slepton interactions has  the form
\bea\mathcal{L}_{\tilde{\mu}\tilde{\tau}H^0}&= &-
\left[(h'_{\mu\tau}\tilde{\mu}_L\rho^{\prime 0}\tilde{\tau}^c_L+
 h'_{\tau\mu}\tilde{\tau}_L\rho^{\prime 0}\tilde{\mu}^c_L
 + h'_{\tau}\tilde{\tau}^c_L\rho^{\prime
 0}\tilde{\tau}_L+  h'_{\mu}\tilde{\mu}^c_L\rho^{\prime 0}\tilde{\mu}_L)
 \right.\crn
  &+&\left. \rho^0(h_{\mu\tau}-h_{\tau\mu}) (\tilde{\nu}^c_{\mu
 L} \tilde{\nu}_{\tau L}-\tilde{\nu}_{\mu L}\tilde{\nu}^c_{\tau
 L}) +  \frac{1}{2}Y_{\tau} \mu_{\rho} \rho^{0*} \tilde{\tau}_L
 \tilde{\tau}^c_L  \right.\crn
 &+&\left. \frac{1}{2} Y_{\mu} \mu_{\rho} \rho^{0*} \tilde{\mu}_L
 \tilde{\mu}^c_L  +  Y_{\nu_{\mu\tau}}\mu_{\rho} \rho^{\prime 0*} \left(\tilde{\nu}^c_{\tau L}
 \tilde{\nu}_{\mu L} -\tilde{\nu}^c_{\mu L}
 \tilde{\nu}_{\tau L}  \right)
  + \mathrm{H.c.}\right] \label{sleptonmass1}\eea
 Vertices of Higgs- slepton-slepton interactions are listed in Table
 \ref{higss-slp-slp1}.
\begin{table}
  \centering
  \begin{tabular}{|c|c|c|c|}
    \hline
    Vertex & Factor & Vertex & Factor \\
    \hline
     $\tilde{\mu}^c_L\tilde{\mu}_L \rho^{\prime 0}$& $-ih'_{\mu}$
      & $\tilde{\tau}^c_L\tilde{\tau}_L\rho^{\prime
 0} $ & $-ih'_{\tau}$\\
     $\tilde{\mu}_L\tilde{\tau}^c_L\rho^{\prime 0}$ & $-ih'_{\mu\tau}$ &
      $\tilde{\tau}_L\tilde{\mu}^c_L\rho^{\prime 0} $  &  $-ih'_{\tau\mu}$\\
 $\rho^0 \tilde{\nu}^c_{\mu
 L} \tilde{\nu}_{\tau L}$ & $-i(h_{\mu\tau}-h_{\tau\mu})$
 & $\rho^0\tilde{\nu}_{\mu L}\tilde{\nu}^c_{\tau
 L}$ &$i(h_{\mu\tau}-h_{\tau\mu})$\\
 $\rho^0\tilde{\tau}^*_L\tilde{\tau}^{c*}_L   $ & $-\frac{i}{2}Y_{\tau} \mu_\rho$
 & $\rho^0\tilde{\mu}^*_L\tilde{\mu}^{c*}_L$ &$-\frac{i}{2}Y_{\mu}\mu_{\rho} $\\
 $\rho^{\prime0}\tilde{\nu}^*_{\tau_L}\tilde{\nu}^{c*}_{\mu_L}   $ & $-iY_{\nu_{\mu\tau}} \mu_\rho$
 & $\rho^{\prime0}\tilde{\nu}^*_{\mu_L}\tilde{\nu}^{c*}_{\tau_L}$ & $ iY_{\nu_{\mu\tau}} \mu_\rho$\\
    \hline
  \end{tabular}
  \caption{ Slepton-slepton-Higgs vertices.}\label{higss-slp-slp1}.
\end{table}
\end{enumerate}
\section{\label{eigenstate}Mass eigenstates
of particles in the  SUSYE331 model}

\subsection{ Neutral Higgs}
The physical states of Higgs (mass eigenstates)
  have been studied in \cite{Dong2}. For
convenience we review the main results in this appendix. First we
expand the neutral Higgs components around the VEVs by
\bea \chi^T&=& \left(%
\begin{array}{ccc}
  \frac{u+S_1+iA_1}{\sqrt{2}}, & \chi^- ,& \frac{w +S_2+i A_2}{\sqrt{2}}  \\
\end{array}%
\right), \hs
\rho^T=\left(%
\begin{array}{ccc}
  \rho^+_1, & \frac{v+S_5+i A_5}{\sqrt{2}} ,&  \rho^+_2 \\
\end{array}%
\right)\crn
\chi^{\prime T}&=& \left(%
\begin{array}{ccc}
  \frac{u'+S_3+iA_3}{\sqrt{2}}, & \chi^{\prime+}
  ,& \frac{w +S_4+i A_4}{\sqrt{2}}  \\
\end{array}%
\right), \hs
\rho^{\prime T}=\left(%
\begin{array}{ccc}
  \rho^{\prime -}_1, & \frac{v'+S_6+i A_6}{\sqrt{2}}
  ,&  \rho^{\prime-}_2 \\
\end{array}%
\right)
 \label{higgs1}\eea
where $\{u,w,u', w', v, v'\}$, $ \{S_1, S_2, S_3, S_4, S_5, S_6\}$
 and $\{ A_1, A_2, A_3, A_4, A_5, A_6\} $ are
VEV, scalar, and pseudo scalar parts of neutral Higgs,
respectively. The Higgs mass spectrum  and the Higgs mass
eigenstates given in \cite{Dong2} showed that:
\begin{description}
    \item Scalar Higgs: Mass eigenstates of six original scalar Higgs
    $ \{S_1, S_2, S_3, S_4, S_5, S_6\}$ are defined as three massless
    eigenstates $\{S^{\prime}_{1a}, S^{\prime}_5, \varphi_{S_{24}}\}$  and
    three massive ones $\{\phi_{S_{24}}, \varphi_{S_{a36}}, \phi_{S_{a36}}\}$.
    The relations between the original  and the mass-eigenstate base
    are
\footnote{ There  are some different definitions for $\gamma$ in
\cite{Dong1,Dong2,Dong3}. In this paper we use notations
identifying with those of \cite{Dong1}.}:
 \bea
\left(%
\begin{array}{c}
  S_1 \\
   S_2 \\
   S_3 \\
   S_4 \\
    S_5 \\
   S_6\\

\end{array}%
\right)&=&
   \left(%
\begin{array}{cccccc}
  c_{\beta}s_{\theta} &-s_{\beta}c_{\theta} &-c_{\beta}
  c_{\theta}&-s_{\alpha}s_{\beta}s_{\theta}
  & -c_{\alpha}s_{\beta}s_{\theta}&0\\
   c_{\beta}c_{\theta}&s_{\beta}s_{\theta}&c_{\beta}
   s_{\theta}& -s_{\alpha}s_{\beta}c_{\theta}
   &-c_{\alpha}s_{\beta}c_{\theta}&0\\
    s_{\beta}s_{\theta}&-c_{\beta}c_{\theta}&s_{\beta}
    c_{\theta}& s_{\alpha}c_{\beta}s_{\theta}
    &c_{\alpha}c_{\beta}s_{\theta}&0\\
   s_{\beta}c_{\theta} &c_{\beta}s_{\theta} &-s_{\beta}
   s_{\theta}&s_{\alpha}c_{\beta}c_{\theta}
  &c_{\alpha}c_{\beta}c_{\theta}&0\\
    0 &0&0&-c_{\alpha}s_{\gamma}&s_{\alpha}s_{\gamma} &c_{\gamma}\\
   0&0&0&c_{\alpha}c_{\gamma} &-s_{\alpha}c_{\gamma} &s_{\gamma}\\
\end{array}%
\right)\left(%
\begin{array}{c}
  S^{\prime}_{1a} \\
   \varphi_{S_{24}} \\
   \phi_{S_{24}} \\
   \varphi_{S_{a36}} \\
    \phi_{S_{a36}} \\
   S^{\prime}_5\\
\end{array}%
\right)
    \label{mscalheigenstate}\eea
where some new notations are defined as  follows:
$$ t_{\theta}\equiv\tan\theta\equiv\frac{u}{w}=\frac{u'}{w'} ,
\; c_{\theta}\equiv\cos\theta, \; s_{\theta}\equiv\sin\theta,$$
$$ t_{\beta}\equiv\tan\beta\equiv\frac{w}{w'},
  \; c_{\beta}\equiv\cos\beta, \; s_{\beta}\equiv\sin\beta,$$
and
$$ t_{\gamma}\equiv\tan\gamma\equiv\frac{v}{v'},\;
s_{\gamma}\equiv\sin\gamma, \; c_{\gamma}\equiv\cos\gamma, $$ and
$\alpha$ is determined through  relations:
$$ \tan{2\alpha}\equiv \frac{-2m^2_{36a}}{m^2_{66a}-m^2_{33a}},
\; c_{\alpha}\equiv\cos\alpha,\; s_{\alpha}\equiv\sin\alpha
$$
$$ m^2_{33a}=\frac{18g^2+g^{\prime2}}{54c^2_{\theta}}
(w^2+w^{\prime2})=\frac{(18g^2+g^{\prime2})w^2}
{54c^2_{\theta}s^2_{\beta}}
$$
$$ m^2_{66a}=\frac{9g^2+2g^{\prime2}}{27}
(v^2+v^{\prime2})=\frac{(9g^2+2g^{\prime2})v^2} {27s^2_{\gamma}}
$$
$$ m^2_{36a}=\frac{9g^2+2g^{\prime2}}{54}
\sqrt{\frac{(v^2+v^{\prime2})(w^2+w^{\prime2})}{c^2_{\theta}}}
=\frac{9g^2+2g^{\prime2}}{54} \frac{v w }{ |c_{\theta}
s_{\gamma}s_{\beta}|}$$
The mass eigenvalues of three physical Higgses  $\phi_{S_{24}},
\varphi_{S_{a36}}$ and  $\phi_{S_{a36}}$ are:
\bea
 m^2_{\phi_{S_{24}}}&=&\frac{g^2}{4}(1+t^2_{\theta})
 (w^2+w^{\prime2})
 =\frac{g^2w^2}{c^2_{\theta}s^2_{\beta}}\\
m^2_{\varphi_{S_{a36}}}&=& \frac{1}{2}\left[ m^2_{33a}+m^2_{66a}
-\sqrt{(m^2_{33a}-m^2_{66a})^2
+4m^4_{36a}}\right]\\
m^2_{\phi_{S_{a36}}}&=& \frac{1}{2}\left[ m^2_{33a}+m^2_{66a}
+\sqrt{(m^2_{33a}-m^2_{66a})^2 +4m^4_{36a}}\right]
\label{masshigss1}\eea
    \item Pseudo-scalar Higgs: There are five Goldstone
    bosons $\{A_5, A_6, A'_1, A'_2,\varphi_A\}$
    and one massive physical Higgs $\varphi_A$. Because the
    $\varphi_A$ does not receive any contributions from  the $A_5, A_6$ and sleptons as well
    as their sneutrinos  only
    couple to $\rho, \rho'$, hence there is no coupling of pseudo
    scalar $\varphi_A$ to muon and tauon at the one loop approximation.
    This is the difference between MSMS and
    our model.

\end{description}
\subsection{ Mass eigenstates of sleptons \label{mass}}

The masses of  sleptons in SUSYE331 models were studied in
details  in \cite{Dong2}. In the work, they assumed that
 lepton numbers are conserved even in the slepton
sector. This assumption leaded to the absence of mixing terms in
slepton sector. Our work is  interested in studying the source of
LFV caused by the mixing between slepton $\tilde{\mu}$ and
$\tilde{\tau}$ and ignore all other sources of FLV . So with two
assumptions of R-parity conversation and the small left-right
mixing in slepton sector, we can base on \cite{Dong2} to write the
mass terms of charged sleptons in the form: \bea
-\mathcal{L}_{\tilde{l}\tilde{l}^*}&=&\sum_{\tilde{l}_{L_a}}
\tilde{m}^2_{\tilde{l}_{L_a}}\tilde{l}^*_{L_a}\tilde{l}_{L_a}+
\left(\tilde{m}^2_{L_{\mu\tau}}\tilde{\mu}^*_L\tilde{\tau}_L+\mathrm{H.c.}\right)
 \crn
&+&\sum_{\tilde{l}_{R_a}}
\tilde{m}^2_{\tilde{l}_{R_a}}\tilde{l}^*_{R_a}\tilde{l}_{R_a}+
\left(\tilde{m}^2_{R_{\mu\tau}}\tilde{\mu}^*_R\tilde{\tau}_R+\mathrm{H.c.}\right)
 \label{massslepton1}\eea where $
\tilde{l}_{L_a}=\{\tilde{e}_L,\tilde{\mu}_L,\tilde{\tau}_L\}$,
$\tilde{l}_{R_a}=\{ \tilde{e}_R,\tilde{\mu}_R,\tilde{\tau}_R\}$
and
\bea \tilde{m}^2_{\tilde{l}_{L_a}} \equiv B_{aa} &=& M^2_{aa}
+\frac{1}{4} \mu_{0a}^2+ \frac{v^{\prime2}}{18}
\lambda^2_{1aa}+\frac{1}{18}\lambda^2_{2a} (u^2+w^2)\crn
&-& \frac{g^2}{2}
\left(H_3-\frac{1}{\sqrt{3}}H_8-\frac{2t^2}{3}H_1\right),\crn
\tilde{m}^2_{\tilde{l}_{R_a}} \equiv C_{aa} &=& m^2_{aa} +
\frac{v^{\prime2}}{18} \lambda^2_{1aa}+g^2t^2H_1,\crn
\tilde{m}^2_{L_{\mu\tau}}\equiv B_{23}&=&  M^2_{23} +\frac{1}{4}
\mu_{02}\mu_{03}+ \frac{1}{18}\lambda_{22}\lambda_{23}
(u^2+w^2),\crn
\tilde{m}^2_{R_{\mu\tau}}\equiv C_{23}&=&  m^2_{23},\crn
 H_1&\equiv& \frac{1}{6} \left[
 (u^2+w^2)\frac{\cos2\beta}{s^2_{\beta}}-
 2v^2\frac{\cos2\gamma}{s^2_\gamma}\right]\crn
 &=& \frac{1}{6} \left[
 (u^2+w^2) \left( \cot^2\beta -1\right)-
 2v^2 \left(\cot^2\gamma -1\right)\right] \crn
 H_3&\equiv&-\frac{1}{4}\left[ u^2\frac{\cos2\beta}{s^2_\beta}-
 2v^2\frac{\cos2\gamma}{s^2_\gamma}\right]\crn
 &=& -\frac{1}{4}\left[ u^2 \left(\cot^2\beta-1\right)-
 2v^2
 \left(\cot^2\gamma-1\right)\right]\crn
H_8&\equiv&-\frac{1}{4\sqrt{3}}\left[
 v^2\frac{\cos2\gamma}{s^2_\gamma}+(u^2-2w^2)
 \frac{\cos2\beta}{s^2_\beta}\right],\crn
 &=& \frac{1}{4\sqrt{3}}\left[
 v^2\left(\cot^2\gamma-1\right)+(u^2-2w^2)
 \left(\cot^2\beta-1\right)\right], \crn
  t^2&\equiv&\left(\frac{g'}{g}\right)^2=\frac{6s^2_W}{3-4s^2_W}
\label{massfactor1}\eea
Note that here we use notations $M^2_{aa}$ and $m^2_{aa}$ in stead
of $m^2_{aL}$ and $m^2_{la}$ in the soft breaking term of
\cite{Dong1}. We assume that the mixing matrix of the
$\tilde{\mu}_{L,R}$ and $\tilde{\tau}_{L,R}$ slepton masses is
given by
\bea-\mathcal{L}_{\tilde{\mu}\tilde{\tau}}&= & \left( \tilde{\mu}_
L^*,\; \tilde{\tau}_L^*
  \right)\left(%
\begin{array}{cc}
  \tilde{m}^2_{\mu_L} & \tilde{m}^2_{L_{\mu\tau}} \\
  \tilde{m}^{*2}_{L_{\mu\tau}} & \tilde{m}^2_{\tau_L} \\
\end{array}%
\right)\left(%
\begin{array}{c}
   \tilde{\mu_L}\\
  \tilde{\tau_L} \\
\end{array}%
\right)
+\left( \tilde{\mu}^{c*}_{ L},\; \tilde{\tau}^{c*}_{L}
  \right)\left(%
\begin{array}{cc}
  \tilde{m}^2_{\mu_R} & \tilde{m}^2_{R_{\mu\tau}} \\
  \tilde{m}^{*2}_{R_{\mu\tau}} & \tilde{m}^2_{\tau_R} \\
\end{array}%
\right)\left(%
\begin{array}{c}
   \tilde{\mu}^c_{ L}\\
  \tilde{\tau}^c_{L} \\
\end{array}%
\right)\crn
  \label{sleptonmass1}\eea
This form  is the same as that given in \cite{Anna1,Anna2} ( for
detail, see \cite{Anna1}, Appendix A.1). Hence, the mass
eigenstates and eigenvalues of sleptons in our model are similar
to that in the MSSM \cite{Anna1,Anna2}.  In particular,  the mass
mixing matrix of the left handed and right handed of sleptons
given in (\ref{sleptonmass1}) produce the mass eigenstates such as
$\{\tilde{l}_{L_2}, \tilde{l}_{L_3}\}$ and $\{ \tilde{l}_{R_2},
  \tilde{l}_{R_3}\}$. The corresponding  mass eigenvalues are
  $\{\tilde{m}^2_{L_2}, \tilde{m}^2_{L_3}\}$ and $\{\tilde{m}^2_{R_2},
  \tilde{m}^2_{R_3}\}$.

 From now we adopt conventions the flavor states of sleptons are
$\tilde{\mu}_L, \tilde{\tau}_L$ and $  \tilde{\mu}^c_L,
\tilde{\tau}^c_L$ while  the mass eigenstates are
$\tilde{l}_{L_2}, \tilde{l}_{L_3}$ and $ \widetilde{l}_{R_2},
\widetilde{l}_{R_3}$, respectively. The relations between these
two kinds of basics are: $ \tilde{\mu}_L= c_L \tilde{l}_{L_2}-s_L
\tilde{l}_{L_3} $, $ \tilde{\tau}_L= s_L \tilde{l}_{L_2}+ c_L
\tilde{l}_{L_3}$, with $c_L=\cos\theta_L$, $s_L=\sin\theta_L$; $
\widetilde{\mu}^c_{L}= c_R \widetilde{l}_{R_2}-s_R
\widetilde{l}_{R_3} $, $ \widetilde{\tau}^c_{L}= s_R
\widetilde{l}_{R_2}+ c_R \widetilde{l}_{R_3}$, with
$c_R=\cos\theta_R$, $s_R=\sin\theta_R$. The mixing parameters
satisfy the following relations: \bea s_L
c_L=\frac{\tilde{m}^2_{L\mu\tau}}{\tilde{m}^2_{L_3}-
\tilde{m}^2_{L_2}},\hs s_R
c_R=\frac{\tilde{m}^2_{R\mu\tau}}{\tilde{m}^2_{R_3}-
\tilde{m}^2_{R_2}}.
 \label{para1}\eea

\subsection{\label{sneum} Sneutrinos}
  The general  Lagrangian which  gains masses for
   sneutrinos is  given in \cite{Dong2} as follows
\bea  \mathcal{L}_{\tilde{\nu}} &=& \left(%
\begin{array}{cc}
  \tilde{\nu}^*_{aL}, & \tilde{\nu}^*_{aR}  \\
\end{array}%
\right) \left(%
\begin{array}{cc}
  A_{ab} & E_{ab} \\
  E_{ab} & G_{ab} \\
\end{array}%
\right) \left(%
\begin{array}{c}
  \tilde{\nu}_{bL} \\
   \tilde{\nu}_{bR}\\
\end{array}%
\right),\label{sneutrino1}\eea
 where
\bea \nu_{aL} \equiv (\nu_{1L},~ \nu_{2L}, ~ \nu_{3L})^T, \hs
\nu_{aR} \equiv \nu^{c*}_{aL} = (\nu^{c*}_{1L}, \nu^{c*}_{2L},
\nu^{c*}_{3L} ), \label{sneutrino3}\eea
 and
 \bea A_{ab} &=& \frac{g^2}{2} \delta_{ab} \left( H_3+ \frac{1}{\sqrt{3}} H_8 -
 \frac{2t^2}{3} H_1\right)+ M^2_{ab}+ \frac{1}{4}
 \mu_{0a}\mu_{0b}, \crn
 &+& \frac{1}{18} \lambda_{2a} \lambda_{2b} (v^2+ w^2)+\frac{2}{9} \lambda_{3ca}
 \lambda_{3cb} v^2, \crn
  G_{ab} &=& -g^2 \delta_{ab} \left(  \frac{1}{\sqrt{3}} H_8 +
 \frac{t^2}{3} H_1\right)+ M^2_{ab}+ \frac{1}{4}
 \mu_{0a}\mu_{0b},  \crn
 &+& \frac{1}{18} \lambda_{2a} \lambda_{2b} (v^2+ u^2)+\frac{2}{9} \lambda_{3ca}
 \lambda_{3cb} v^2, \crn
 E_{ab}&=& -\frac{1}{\sqrt{2}}\left[ (\varepsilon_{ab}- \varepsilon_{ba} )
 v + \frac{1}{6}
 \mu_{\rho} v' (\lambda_{3ab}-\lambda_{3ba})\right] .
  \label{sneutrino2}\eea

If  the LFV happens only in the $\{ \tilde{\nu}_{\mu},
\tilde{\nu}_{\tau}\}$ sector,  we can rewrite the  non-vanishing
terms given in (\ref{sneutrino2}) in more explicit formulas:
\bea m^2_{\tilde{\nu}_{aL}} \equiv A_{aa} &=& \frac{g^2}{2}
 \left( H_3+ \frac{1}{\sqrt{3}} H_8 -
 \frac{2t^2}{3} H_1\right)+ M^2_{aa}+ \frac{1}{4}\mu^2_{0a} \crn
 &+&
 \frac{1}{18} \lambda^2_{2a} (v^2+ w^2)+\frac{2}{9}
 v^2 \sum_{c}\lambda^{2}_{3ca}, \crn
m^2_{\tilde{\nu}_{aR}} \equiv G_{aa} &=& -g^2 \left(
\frac{1}{\sqrt{3}} H_8 +
 \frac{t^2}{3} H_1\right)+ M^2_{aa}+ \frac{1}{4}
 \mu^2_{0a} \crn
 &+& \frac{1}{18} \lambda^2_{2a}  (v^2+ u^2)+\frac{2}{9}v^2
 \sum_{c}\lambda^{ 2}_{3ca}, \crn
 m^2_{\tilde{\nu}_{L\mu\tau}} \equiv A_{23} &=&  M^2_{23}+
 \frac{1}{4}\mu_{02} \mu_{03}+
 \frac{1}{18} \lambda_{22} \lambda_{23} (v^2+ w^2)+\frac{2}{9}  v^2
 \lambda^{2}_{3c2}\lambda_{3c3},\crn
  m^2_{\tilde{\nu}_{R\mu\tau}} \equiv A_{23} &=&  M^2_{23}+
 \frac{1}{4}\mu_{02} \mu_{03}+
 \frac{1}{18} \lambda_{22} \lambda_{23} (v^2+ u^2)+\frac{2}{9}  v^2
 \lambda_{3c2}\lambda_{3c3}.\label{sneutrino4}\eea
 Similar to the charged sleptons sector, we  denote by $\{ \tilde{\nu}_{\mu_L},~
 \tilde{\nu}_{\tau_ L}, ~\tilde{\nu}_{\mu_R},
 ~ \tilde{\nu}_{\tau_R}\}$ the  flavor
 eigenstates while by $\{ \tilde{\nu}_{L2},~ \tilde{\nu}_{L3},
  ~\tilde{\nu}_{ R2},~ \tilde{\nu}_{R3}\}$ the mass eigenstates.
  Also, notations  $\{ \tilde{m}^2_{\nu_{L2}},~ \tilde{m}^2_{\nu_{L3}},$ $~
  \tilde{m}^2_{\nu_{R2}}, \tilde{m}^2_{\nu_{R3}}\}$ denote the  mass
  eigenstates of sneutrinos.  Here  the relations between two
  bases are:
\bea   \widetilde{\nu}_{\mu_L} &=& c_{\nu_L}
\tilde{\nu}_{L2}-s_{\nu_L} \tilde{\nu}_{L3}, \hs
\widetilde{\nu}_{\tau_L} = s_{\nu_L} \tilde{\nu}_{L2}+ c_{\nu_L}
\tilde{\nu}_{L3}, \crn \widetilde{\nu}_{\mu_R} &=& c_{\nu_R}
\tilde{\nu}_{R2}-s_{\nu_R} \tilde{\nu}_{R3}, \hs
\widetilde{\nu}_{\tau_R} = s_{\nu_R} \tilde{\nu}_{R2}+ c_{\nu_R}
\tilde{\nu}_{R3}, \crn s_{\nu_L}
c_{\nu_L}&=&\frac{\tilde{m}^2_{\nu_{L\mu\tau}}}{\tilde{m}^2_{\nu_{L3}}-
\tilde{m}^2_{\nu_{L2}}},\hs s_{\nu_R}
c_{\nu_R}=\frac{\tilde{m}^2_{\nu_{R\mu\tau}}}{\tilde{m}^2_{\nu_{R3}}-
\tilde{m}^2_{\nu_{R2}}}. \label{sneutrino5} \eea


\begin{thebibliography}{999}

\bibitem{pdg} K. Nakamura {\it et al.} [Particle Data Group], J. Phys. G
\textbf{37}, 075021 (2010).

\bibitem{331m} F. Pisano and V. Pleitez, Phys. Rev.  D {\bf 46}, 410 (1992);
P. H. Frampton, Phys. Rev. Lett. {\bf 69}, 2889 (1992); R. Foot,
O. F. Hernandez, F. Pisano and V. Pleitez, Phys. Rev. D {\bf 47},
4158 (1993).

\bibitem{331r} M. Singer, J. W. F. Valle and J. Schechter, Phys.
Rev. D {\bf 22}, 738 (1980);  R. Foot, H. N. Long and Tuan A.
Tran, Phys. Rev. D {\bf 50}, 34(R) (1994) [arXiv: hep-ph/9402243]
; J. C. Montero, F. Pisano and V. Pleitez, Phys. Rev. D {\bf 47},
2918 (1993); H. N. Long, Phys. Rev. D {\bf 54}, 4691 (1996); Phys.
Rev. D {\bf 53}, 437 (1996); H. N. Long,  {\it Mod. Phys. Lett.}
{\bf A13}, (1998) 1865.


\bibitem{longvan} H. N. Long and V. T. Van, J. Phys. G {\bf 25}, 2319
(1999).

\bibitem{ecq}F. Pisano, Mod. Phys. Lett A {\bf 11}, 2639 (1996);
A. Doff and F. Pisano, Mod. Phys. Lett. A {\bf 14}, 1133 (1999);
C. A. de S. Pires and O. P. Ravinez, Phys. Rev. D {\bf 58}, 035008
(1998); C. A. de S. Pires, Phys. Rev. D {\bf 60}, 075013 (1999);
P. V. Dong and H. N. Long, Int. J. Mod. Phys. A {\bf 21}, 6677
(2006).

\bibitem{ecn331r} P. V. Dong, H. N. Long, D. T. Nhung, D.
V. Soa, Phys. Rev. D 73, 035004 (2006), P. V. Dong, H. N. Long,
Adv.  High Energy Phys., {\bf 2008}, 739492 (2008); arXiv:
0804.3239; W. Ponce, Y. Giraldo, L. A. Sanchez, Phys. Rev. D 67,
075001 (2003).

\bibitem{LHCAtlas} ATLAS Collaboration (G. Aad et al.), January
2009.
\bibitem{ILC}ILC Collaboration (Gerald Aarons (SLAC) et al.).
September 2007.

\bibitem{thamkhao}E. Arganda, A. M. Curiel, M. J.  Herrero, D. Temes, Phys. Rev. D \textbf{71} (2005) 035011,
arxiv: hep-ph/0407302.
\bibitem{Anna1} A. Brignoble,  A.  Rossi, Phys. Lett. \textbf{B 66} (2003)
217, arXiv:hep-ph/0304081.
\bibitem{Anna2} A. Brignole, A. Rossi, Nucl. Phys.  \textbf{B 701}
(2004),3-53, arXiv:hep-ph/0404211.
\bibitem{cruz}J. L. Diaz-Cruz, D. K. Gosh and S. Morreti, Phys. Lett. \textbf{B 679} (2009)
376, arXiv:0809.5158.
\bibitem{black}  G.  Blankenburg, J.  Ellis and G. Isidori,
{\it Flavour-Changing Decays of a 125 GeV Higgs-like Particle},
e-Print: arXiv:1202.5704 [hep-ph]
\bibitem{Koji} S. Kanemura, K. Matsuda, T. Ota, T. Shindou, E. Takasugi and K.Tsumura
Phys. Lett. \textbf{B}599(2004)83, hep-ph/0406316; S. Kanemura, T.
Ota, T. Shindou and K.Tsumura, Phys. Lett. \textbf{D}73 (2006)
016006, hep-ph/0505191.

\bibitem{Babu1} K. S. Babu and C. Kolda, Phys. Lett \textbf{B 451} (1999) 77,
arxiv: 9811308.
\bibitem{Babu} K. S. Babu and C. Kolda, Phys. Rev. Lett. \textbf{89} (2002)
241802, arXiv: 0206310; K. S. Babu and C. F. Kolda, Phys. Rev.
Lett. \textbf{84} (2000) 228, e-Print: hep-ph/9909476.
\bibitem{Guasch} J. Guasch, W. Hollik and S. Penaranda, Phys. Lett.
\textbf{B 515} (2001) 367, arxiv: 0106027.
\bibitem{Cannoni} M. Cannoni and O. Panella, Phys. Rev. D \textbf{79} (2009)
056001, Arxiv: 0812.2875;  Phys. Rev. D 81 (2010) 036009,
arXiv:0910.3316; Nuovo Cim. \textbf{C 33} (2010) 183,
arXiv:1002.3697.
\bibitem{Carena} M. S. Carena, D. Garcia, U. Nierste and C. E. M.
Wagner, Nucl. Phys. \textbf{B 577} (2000) 88, arXiv: 9912516.
\bibitem{Goto} T. Goto, Y. Okada and Y. Yamamota, Phys. Rev. D \textbf{83} (2011)
053011, arXiv:1012.4385.
\bibitem{Tom} T. Blazek and S. Raby, Phys. Rev. D \textbf{59} (1999) 095002,
Arxiv: 9712257.
\bibitem{Gino} G. Isidori and A. Retico, JHEP \textbf{0111} (2001) 001, Arxiv: 0110121.
\bibitem{Dong1}
P. V. Dong, D. T. Huong, M. C. Rodriguez, H.  N. Long, Nucl. Phys.
\textbf{B 772} (2007) 150; e-Print: hep-ph/0701137.
\bibitem{Dong2}P. V. Dong, Tr. T. Huong, N. T. Thuy and H. N.
Long, JHEP \textbf{0711} (2007)073, arXiv:0708.3155 [hep-ph].
\bibitem{Dong3}P. V. Dong, D. T. Huong, N. T. Thuy and H. N.
Long, Nucl. Phys. \textbf{B 795} (2008) 361; arXiv:0707.3712
\bibitem{Long2}D. T. Huong and H. N. Long, JHEP
\textbf{0807} (2008) 049, arXiv:0804.3875.

\bibitem{exp} D. Atwood, L. Reina and A. Soni, Phys. Rev D \textbf{55}, 3156 (1997)
\bibitem{paradisi1}P. Paradisi, JHEP \textbf{0608} (2006) 047, e-Print:
hep-ph/0601100.
\bibitem{lorenzo} J. L. Diaz-Cruz, JHEP \textbf{0305} (2003) 036, e-Print:
hep-ph/0207030.
\bibitem{okada1} Y. Okada, K. Okumura, Y. Shimizu, Phys.Rev. D \textbf{61} (2000)
094001, e-Print: hep-ph/9906446.
\bibitem{belle1}Y. Miyazaki \emph{et al}. [Belle Collaboration], Phys. Lett. \textbf{B 660} (2008)
154 ,arXiv:0711.2189.
\bibitem{hisano}J.Hisano and D.
Nomura, Phys. Rev. D \textbf{59}, 116005 (1999);

R. Bernstein, in: 4th International Workshop on Nuclear and
Particle
 Physics at J-PARC (NP08),Mito, Ibaraki,
 Japan, March
 2008,http://j-parc.jp/researcher/Hadron/NP08/index.html.
\bibitem{s331r} J. C. Montero, V. Pleitez, M. C. Rodriguez,
Phys. Rev. D {\bf 70} (2004) 075004.

\end{thebibliography}
\end{document}